\documentclass[12pt,a4paper]{article}
\pdfoutput=1
\usepackage{cite}
\usepackage{jheppub}
\usepackage{amsmath,amsthm,amssymb,graphicx,slashed}
\usepackage{booktabs}
\usepackage{hyperref}
\usepackage[usenames,dvipsnames]{xcolor}
\usepackage{cleveref}
\usepackage[multiple]{footmisc}

\numberwithin{equation}{section}

\def\be{\begin{equation}}
\def\ee{\end{equation}}
\def\ba{\begin{align}}
\def\ea{\end{align}}
\def\beq{\begin{eqnarray}}
\def\eeq{\end{eqnarray}}

\title{Ineffective Higher Derivative Black Hole Hair}

\author{Kevin Goldstein,}
\author{James Junior Mashiyane}

\affiliation{Mandelstam Institute for Theoretical Physics, School of Physics, and National Institute for Theoretical Physics, University of the Witwatersrand, Johannesburg, WITS 2050, South Africa}

\emailAdd{kevin.goldstein@wits.ac.za}
\emailAdd{James.Mashiyane@students.wits.ac.za}

\abstract{
	Inspired by possibility that the Schwarzschild black hole may not be the unique spherically symmetric vacuum solution to generalisations of general relativity, we consider black holes in pure fourth order higher derivative gravity treated as an effective theory. Such solutions may be of interest in addressing the issue of higher derivative  hair or during the later stages of black hole evaporation. Non-Schwarzschild solutions have been studied but we have put earlier results on a firmer footing by finding a systematic asymptotic expansion for the black holes and matching them with known numerical solutions obtained by integrating out from the near horizon region. These asymptotic expansions can be cast in the form of trans-series expansions which we conjecture will be a generic feature of non-Schwarzschild higher derivative black holes.   Excitingly we find a new branch of solutions with lower free energy than the Schwarzschild solution, but as found in earlier work, solutions only seem to exist for black holes with large curvatures meaning that one should not generically neglect even higher derivative corrections. This suggests that one effectively recovers the non-hair theorems in this context.  
}

\begin{document}

\maketitle

\section{Introduction}\label{sec:intro}

   Birkoff's theorem~\cite{jebsen1921general, birkhoff1923relativity, ANDP:ANDP19233771804, eiesland1925group}  implies  that the unique spherically symmetric, asymptotically flat, vacuum solution of general relativity is the Schwarzschild black hole. Despite apparently having a large entropy, this black hole is  completely characterised   by just one parameter -- namely its mass. Essentially taking $R_{\mu\nu}=0$, together with spherical symmetry and asymptotic flatness, gives 
  \begin{equation}
  \label{eq:Schw}
  ds^2=-\left({1-\frac{2M}{r}}\right)dt^2+\frac{dr^2}{\left(1-\frac{2M}{r}\right)}+r^2 d\theta^2+r^2\sin^2\theta d\phi^2~.
  \end{equation}
  
  Recently, it was shown in~\cite{Lu:2015cqa} that in fourth order gravity, the stronger assumption of a static spherically asymptotically flat solution leads to a weaker restriction that the Ricci scalar, $R$, but {\em not} the Ricci tensor, $R_{\mu\nu}$, is zero\footnote{It is not hard to see that in general relativity vacuum solutions necessarily have $R=R_{\mu\nu}=0$.}\footnote{For some interesting earlier work on Birkoff's theorem in higher derivative gravity see \cite{Oliva:2011xu,Oliva:2012zs}}. This in turn opens up the exciting possibility that there are so-called non-Schwarzschild black holes. For reasons we will review later, the fourth order equations of motion are numerically unstable (or ``stiff'') and one can only integrate them for a while before they diverge. In \cite{Lu:2015cqa} it was found that  by  tuning the initial conditions near the horizon one can successively integrate out further and further approaching asymptotically flat space. This is convincing evidence for the existence of such solutions.   We have put these results on a firmer footing by matching a systematic asymptotic expansion to the numerical solutions. While it was known that these solutions were characterised by just two near horizon parameters, from the asymptotic form of the solution, we explicitly see that in addition to the mass, the solutions are characterised by the size of  a massive Yukawa mode associated with the higher derivative terms. Armed with the asymptotic form of the solution we were also able to find a new branch of solutions that was previously missed.
  
  We are interested in studying these solutions for three reasons. 
  Firstly, the fact that Birkhoff's theorem is weakened once we add higher derivatives, suggests the possibility of higher derivative hair which could play a role in understanding black hole entropy. Although one finds that these solutions are only characterised by one additional parameter, it does suggest that even higher order terms in the Lagrangian could lead to more hair. 
  Secondly, as a Schwarzschild black hole shrinks by Hawking  evaporation, treating gravity as an effective theory, one expects that at some length scale, higher derivative terms may become important. This means that non-Schwarzschild black holes may play some role in the late time evolution of black holes. 
  Finally, we would like to study asymptotically AdS non-Schwarzschild black holes which could have interesting holographic interpretations. This paper serves as a warm up for that project.  
   
  The laws of black hole thermodynamics \cite{bh1,haw} generalise once one adds higher derivative terms \cite{Wald:1993nt, Jacobson:1993vj, Iyer:1994ys,Jacobson:1994qe}. In particular the Bekenstein-Hawking entropy,  given by 
  \begin{equation}
  \label{BH}
  S_\text{BH}=\tfrac{1}{4}A~,
  \end{equation}
  where $A$ is the horizon area (in Plank units), 
  is replaced by the Wald entropy which, for a spherical solution, is given by \cite{Sen:2007qy}
  \begin{equation}
  \label{Wald}
  S_\text{Wald}=-8\pi\int_{\text{Horizon}}d\theta d\phi\frac{\delta \mathcal{S}}{\delta R_{rtrt}}\sqrt{-g_{tt} g_{rr}}~,
  \end{equation}
  with $\mathcal{S}$ the relevant action. We take the non-Schwarzschild black hole's entropy to given by the Wald entropy. 
  
  Although we will discuss it in detail later (see \Cref{fig:Tr0,fig:Mr0,fig:ST,fig:FT}  for detailed plots), it is worth mentioning some features of  the thermodynamics of the non-Schwarzschild black holes. Unlike the Schwarzschild black hole, the horizon radius decreases as these black holes get colder. At a fixed horizon size, one has both a Schwarzschild and a non-Schwarzschild solution except for one critical point where the solutions coincide. Below this critical size, the non-Schwarzschild black hole is colder than the Schwarzschild solution, while above the critical size it is hotter than the Schwarzschild solution. We will call the two sets of solutions the cold and hot branches respectively\footnote{Only the hot branch was found in \cite{Lu:2015cqa}.}.
  Like the  Schwarzschild black hole their entropy decreases as one increases the temperature. However, on the cold branch, the black hole's entropy seems to approach a finite value as the temperature approaches zero. Bizarrely on the hot branch, as one increases the temperature, the entropy reaches zero at some point and then becomes negative as the temperature increases further. We cannot see how this makes any physical sense. Finally it is interesting to note that, at a fixed temperature, both branches have lower entropy than the Schwarzschild solution. Intriguingly, also at fixed temperature, the cold branch has a lower free energy,  while the hot branch has a higher free energy, than the corresponding Schwarzschild one.
  
  In addition to troubling thermodynamics, both branches of the non-Schwarzschild solutions are highly curved which would not generically be consistent with neglecting even higher order curvature terms in the action. One could insist on truncating the theory rather than treating it as an effective theory but this will come at the cost of introducing ghosts \cite{stelle1978classical}. These unphysical excitations could be responsible for some of the strange thermodynamic features of the solutions. This suggests that the only physically reasonable black hole one can take is the Schwarzschild solution eliminating the prospect of higher derivative hair at the first non-trivial order. We have called this class of hair ``ineffective'' since it disappears if one insists on the self-consistency of the low energy effective description\footnote{In principle, there could be special situations in which it makes sense to consider a truncated theory in which case our objections to this hair falls way. Unfortunately it does not seem likely that such ``special'' hair could generically play a role in understanding black hole entropy.}.
  
  On a technical note, the asymptotic expansion we find takes the form of a trans-series. We conjecture that this will be a generic feature and may prove to be a useful tool for studying the asymptotics of solutions to higher derivative theories in general.  
  
  In \autoref{sec:NS} we review results on non-Schwarzschild black holes. Then in \autoref{sec:asymp}, we show how to systematically find the asymptotic form of the metric followed by a discussion of the thermodynamics  one extracts by matching the numerical and asymptotic solutions  in \autoref{sec:match}. In \autoref{sec:conc} we present our conclusions and prospects for future work. Finally some technical aspects are left to \autoref{ap:asymp} and  \autoref{ap:num} which cover further details of  the asymptotic expansion and our numerical results respectively.

\section{Non-Schwarzschild black holes}\label{sec:NS}

We review some of the results of \cite{Lu:2015cqa,Lu:2015psa} on 
non-Schwarzschild black holes which will be useful for us (with some added details, and slight changes in notation and emphasis). We consider a general pure gravitational action of the form
\begin{equation}
\label{eq:Action}
\mathcal{S}	=\frac{1}{8\pi L_P^2}\int d^4x\sqrt{-g}\; \left( R +L_P^2\left(\beta  R^2 -\alpha  C^{\alpha\beta\mu\nu} C_{\alpha\beta\mu\nu} \right) +\mathcal{O}\left(\frac{L_P^3}{L_C^3}\right)+\ldots\right)~,
\end{equation}
where  $C^{\alpha\beta\mu\nu}$ is the Weyl tensor, and $\alpha$ and $\beta$ are dimensionless constants\footnote{One can add a Gauss-Bonnet term which, in four dimensions, does not affect the equations of motion, but can shift the Wald entropy.}. $L_P$ is the Planck scale and $L_C$ is some length scale associated with the curvature scale of our solution at the horizon -- for a Schwarzschild black hole $L_C\sim M^{-1}$. Up to $\mathcal{O}\left(\frac{L_P^2}{L_C^2}\right)$, the equations of motion, $E_{\mu\nu}$ , which follow from \eqref{eq:Action} are \cite{Lu:2015cqa}:
\begin{eqnarray}
\label{eq:eom}
E_{\mu\nu}=0&=& R_{\mu\nu}-\tfrac{1}{2}g_{\mu\nu}R  
-4\alpha L_P^2\left(\nabla^\alpha\nabla^\beta +\tfrac{1}{2} R^{\alpha\beta}\right)C_{\mu\alpha\nu\beta}\cr
&& +2\beta L_P^2\left(R\left( R_{\mu\nu}-\tfrac{1}{4}g_{\mu\nu}R\right)+(g_{\mu\nu}\nabla^\alpha\nabla_\alpha-\nabla_\mu\nabla_\nu)R\right)~,
\end{eqnarray}
where $\nabla_\alpha$ is the covariant derivative. One can check that vacuum solutions of general relativity satisfy \eqref{eq:eom} so that in particular the Schwarzschild blackhole is a solution\footnote{The non-trivial result one has to use is the Bianchi identity for the Weyl Tensor~\cite{hawkingellis}, $\nabla^\rho C_{\rho\sigma\alpha\beta}=\nabla_{[\alpha} R_{\beta]\sigma}+\frac{1}{6}g_{\sigma[\alpha}\nabla_{\beta]}R$, so that $R_{\mu\nu}=R=0\Rightarrow \nabla^\alpha C_{\mu\alpha\nu\beta}=0$.}.

In  \cite{stelle1978classical}, it was shown that when one studies linear fluctuations of \eqref{eq:eom} about flat space, in addition
to the usual massless spin-2 graviton mode, there is are massive spin-0 (with $m_0^2=1/(6\beta L_P^2)$ ) and spin-2 (with $m_2^2=1/(2\alpha L_P^2)$) modes.  
The massive modes will introduce Yukawa like behaviour, $e^{\pm m r}/r$, in the asymptotic form of the metric. This means that, to obtain asymptotically flat solutions, one needs to  finely tune conditions in the bulk of the spacetime to avoid the growing Yukawa mode. Furthermore, the growing mode makes the numerical integration of the equations tricky since rounding errors can inadvertently  excite the growing mode leading to so-called ``stiff equations'' which easily blow up\footnote{Of course such Yukawa modes could also have implications for the stability of such solutions against perturbations which we do not consider at this stage.}.  
Also notice that one needs to take $\alpha$ and $\beta$ positive to avoid tachyon-like modes. 

In \cite{Lu:2015cqa} it was shown that any asymptotically flat black hole solution to \eqref{eq:eom} must have $R=0$ but not necessarily $R_{\mu\nu}=0$. As is well known \cite{jebsen1921general, birkhoff1923relativity, ANDP:ANDP19233771804, eiesland1925group},  spherical symmetry and $R_{\mu\nu}=0$ imply that the only black hole solution is the Schwarzschild one. With the weaker requirement that $R=0$, there is the possibility of other solutions. Given the complexity of \eqref{eq:eom}, it is challenging to find analytic solutions but one can resort to numerics and perturbative expansions. The matching of a near horizon expansion with numerical integration was studied in \cite{Lu:2015cqa} -- we have further studied an asymptotic expansion about flat space which complements this earlier work. 

Following on from the discussion in the previous paragraph, we will only be interested in solutions with $R=0$ from now on. This means that the value of $\beta$ will not affect the equation of motion so we will ignore it\footnote{ \cite{Alvarez-Gaume:2015rwa} investigates  non-Schwarzschild   higher derivative black holes with $\alpha=0$. They are working in quadratic gravity which does not start with the usual Einstein-Hilbert term $R$.   Since the equations of motion are different in that case, our argument does not apply.}. Furthermore, once we neglect the terms involving $\beta$, it is not hard to see that the trace of the equations of motion \eqref{eq:eom} leads to $R=0$.    
 
\subsection{Near horizon expansion}
\label{sec:NH}
We now consider the near horizon form of the metric which gives the boundary conditions for numerical integration of the equations of motion. Given that we can neglect the terms involving $\beta$, the only length scale in \eqref{eq:eom} is $L_\alpha=\sqrt{2\alpha} L_P$.\footnote{We put in a factor of $\sqrt{2}$ into the definition of $L_\alpha$ for later convenience. }  It is convenient to scale out this dependence by defining a  dimensionless radial coordinate $\rho=m_2 r=\frac{r}{L_\alpha }$.  We take the spherically symmetric ansatz
\begin{equation}
\label{anz1}
ds^2= -h(\rho)dt^2+{L_\alpha^{2}}\frac{d\rho^2}{f(\rho)}+{{L_\alpha^{2}}}{\rho^2}\left(d\theta^2+\sin^2\theta d\phi^2\right)~.
\end{equation}
With this ansatz, 
 the trace of the equations of motion, $g^{\mu\nu}E_{\mu\nu}=0$, \footnote{As mentioned in the beginning of this section, the trace of the equations of motion, reduces to $R=0$ in this case.} gives
%%%%%%%%%%
\begin{equation}
\label{eq:ETr}
4h^2( f-1) +4\rho h  \left( h f\right)'+\rho ^2 \left(h \left(f'
h'+2 f h''\right)-f h'^2\right)=0~,
\end{equation}
where, $'$, denotes derivatives with respect to $\rho$, and the $(\rho\rho)$  equation of motion, $E_{\rho\rho}=0$,  gives
%%%%%%%%%%
\begin{equation}
\label{eq:Err}
\begin{array}{l}
0=\cr
16 h^4(1-f^2 ) +32 \rho  f^2 h^3 h' 
+4\rho ^2 h^2\left(2fh  \left(2 h \left(f''+3\right)- f' h'\right)-  h^2 \left( f'^2+12\right)
-f^2 \left(8 h h''+  h'^2\right)\right) \\
+4\rho ^3 \left( h^3 f'^2 h'+4 f h \left( h f' h'^2-h^2 \left( \left(f''-3\right) h'+ f' h''\right)\right)-f^2  \left(4 h''' h^2+5 
h'^3-12 h h' h''\right)\right)\\
+\rho ^4 \left(2fhh' \left(2h \left(f'' h'+2 f'  h''\right)-3  f' {h'}^2\right)-h^2 f'^2 h'^2+f^2 \left(7 h'^4-12 h h'^2 h''+4h^2 \left(2 h''' h'- h''^2\right)\right)\right)~.
\end{array}
\end{equation}
%%%%%%%%%

%%%%%%%%%

To study black hole solutions, one takes the following expansion about the horizon (at some $\rho_0$) :
\begin{eqnarray}
h&=&c\left[(\rho-\rho_0)+h_2(\rho-\rho_0)^2+h_3(\rho-\rho_0)^3+\ldots\right]~,\label{anz2}\\ 
f&=&f_1(\rho-\rho_0)+f_2(\rho-\rho_0)^2+f_3(\rho-\rho_0)^3+\ldots~.\label{anz3}
\end{eqnarray}
Note that the constant $c$ can be shifted by a trivial rescaling of $t$ and consequently does not enter into the equations of motion (although	 $c$ does need to be tuned if one demands that $h$ asymptote to $1$ at the end). One can show that $\rho_0=1/f_1$ corresponds to the Schwarzschild solution so that the parameter \[\delta=f_1\rho_0-1~,\] gives an indication of the degree to which our solution deviates from the Schwarzschild case. 

 Now using \eqref{anz2}-\eqref{anz3} and evaluating the periodicity of  Euclidean time at the horizon, one finds the temperature to be
 %%%%%%%%%%
 \begin{equation}
 \label{temp}
 T= \frac{1}{2\pi L_\alpha }\sqrt{c f_1}~, 
 \end{equation}
 %%%%%%%%% 
 and from \eqref{Wald}, one finds the Wald entropy: 
 \begin{eqnarray}
 \label{eq:Wald2}
 S_\text{Wald}&=& S_\text{BH}-(2\pi\delta) L_\alpha^2 ~,
 \end{eqnarray}
where as usual the Bekenstein-Hawking entropy, $S_\text{BH}$, is one quarter the horizon area. It will turn out, for a given mass, the Schwarzschild and non-Schwarzschild black holes have a different horizon area (see \autoref{fig:Mr0}). This means that when determining entropy as a function of mass, both terms in \eqref{eq:Wald2} lead to shifts in the entropy. 

Systematically solving \eqref{eq:ETr}-\eqref{eq:Err} for  $f_i$ and $h_i$ at each order in $(\rho-\rho_0)^i$, one finds that there are two free parameters which can be taken to be $\delta$ and $\rho_0$. For instance at second order one finds\footnote{Given the system's numerical instability, we solved the equations up to $18^\text{th}$ order to accurately set up the initial conditions.}
\begin{eqnarray}
f_2=-\frac{1+(3-\tfrac{3}{4}\rho_0^2)\delta+2\delta^2}{(1+\delta^2)\rho_0^2}~,\qquad
h_2=-\frac{1+(3+\tfrac{1}{4}\rho_0^2)\delta+2\delta^2}{(1+\delta^2)\rho_0^2}~.
\end{eqnarray}

%\section{Numerical Integration}\label{sec:NI}

\section{Asymptotic expansion}\label{sec:asymp}

To give more solid evidence for non-Schwarzschild black holes we will find a systematic asymptotic expansion which we can match to the near horizon expansion discussed in the previous section. It appears that, in principle, this expansion is extendable to any order -- albeit at the cost of having to evaluate some  integrals beyond second order numerically. 
  
Generalising 
\cite{stelle1978classical}, it turns out to be more convenient to rewrite the metric as follows
\begin{equation}
\label{eq:metric2}
ds^2= -h(\rho)dt^2+{L^{2}_\alpha}A(\rho)d\rho^2+{L^{2}_\alpha}{\rho^2}\left(d\theta^2+\sin^2\theta d\phi^2\right) ~,
	\end{equation} 
	where as before $\rho=m_2 r$.
We then consider an expansion about flat space
\begin{eqnarray}
\label{eq:expansion}
h&=& 1 + \epsilon h_{(1)}  + \epsilon^2 h_{(2)}+\ldots~, \cr
A&=& 1 + \epsilon A_{(1)}+ \epsilon^2 A_{(2)}+\ldots~, 
\end{eqnarray}
together with the two following linear combinations of the equations of motion \eqref{eq:eom}:
\begin{eqnarray}
g^{\mu\nu}E_{\mu\nu}&=&0  \label{eq:EOM1}~,\\
|g^{00}| E_{00}+g^{ij}E_{ij}  &=&0 \qquad \text{where }~ i,j \in \{1,2,3\}~.\label{eq:EOM2}
\end{eqnarray}
Recall from the previous section that \eqref{eq:EOM1} implies $R=0$ in this context.
\subsection{First order expansion}
Substituting \eqref{eq:expansion} into \eqref{eq:EOM1} and \eqref{eq:EOM2}  and expanding to order $\epsilon$, one finds
\begin{eqnarray}
2 \rho A_{\text{(1)}}'
+2 A_{\text{(1)}}
-\rho^2 h_{\text{(1)}}''
-2 \rho h_{\text{(1)}}'&=&0~,  \label{eq:EOM1B}\\
2 \rho ^3 A_{\text{(1)}}'''+2 \rho ^2 A_{\text{(1)}}''-4 \rho  A_{\text{(1)}}'+4 A_{\text{(1)}}&&\cr-3 \rho ^4 h_{\text{(1)}}''+2 \rho ^4
   h_{\text{(1)}}''''-6 \rho ^3 h_{\text{(1)}}'+8 \rho ^3 h_{\text{(1)}}'''&=&0~.\label{eq:EOM2B}
\end{eqnarray}
These two equations can be decoupled and easily integrated using new variables. Letting
\begin{eqnarray}
Y_{(1)} &=& \frac{1}{\rho}(\rho A_{(1)})'~, \label{eq:subs1}\\
Z_{(1)} &=& (\rho^2 h_{(1)}')'\label{eq:subs2}~,
\end{eqnarray}
one finds that \eqref{eq:EOM1B} can be written
\begin{eqnarray}
%Y_{(1)} &=& \frac{1}{r}(r A_{(1)})'~,\label{eq:subs1}\\
2\rho Y_{(1)} &=& Z_{(1)}\label{eq:EOM1C}~.
\end{eqnarray}
Then using our new variables and \eqref{eq:EOM1C} to eliminate $Z_{(1)}$, \eqref{eq:EOM2B} becomes
\begin{equation}
\label{eq:Y1}
Y_{(1)} - Y_{(1)}''=0~.
\end{equation}
Using \eqref{eq:subs1}, in terms of $A_{\text{(1)}}$ , \eqref{eq:Y1},  becomes
\begin{equation}
\label{eq:A1}
\mathcal{D} A_{\text{(1)}} =0~,
\end{equation}
where
\begin{equation}
\label{eq:A1b}
\mathcal{D}= \left(\frac{1}{\rho }-\frac{2}{\rho ^3}\right)+\left(\frac{2}{\rho ^2}+1\right) \frac{d}{d\rho}-\frac{1}{\rho }\frac{d^2}{d\rho^2}-\frac{d^3}{d\rho^3}~.
\end{equation}
Solving \eqref{eq:A1} we find\footnote{Multiplying \eqref{eq:A1} by $\rho$ one finds that $(\rho \mathcal{D}) A$ can be written as a total derivative 
	$\left(\left(\frac{2}{\rho}+\rho\right)A - \rho A''  \right)'$. This is easily integrated leading to  a second order linear equation which can then be integrated to give \eqref{eq:A1sol}. }
%\footnote{Guessing that a particular solution to \eqref{eq:A1} is of the same form of the Schwarzschild solution, we make the ansatz, $A_{(1)}= \frac{1}{\rho} \mathcal{A}(\rho) $, hoping to reduce the order of the equation. Indeed we find \eqref{eq:A1} becomes, $\rho^2 u'' -2\rho u' +(2-\rho^2)u=0$ where $u=\mathcal{A}'$. If not for the $\rho^2 u$ term, the preceding equation would be Euler's equation with the solutions $u= \alpha_1 \rho + \alpha_2 \rho^2$. This in turn suggests making the substitution  $u = \rho v $ and  we find, $v''-v=0$, which is easily solved and finally working backwards we obtain \eqref{eq:A1sol}.}
\begin{equation}
\label{eq:A1sol}
A_{\text{(1)}}=\frac{c_1}{\rho }+\frac{c_2 e^{-\rho } (-\rho -1)}{\rho }+\frac{c_3 e^{\rho } (\rho -1)}{2 \rho }~,
\end{equation}
where $c_1$, $c_2$ and $c_3$ are  integration constants. Substituting \eqref{eq:A1sol} this into  \eqref{eq:EOM1C} (using \eqref{eq:subs1} \& \eqref{eq:subs2}) results in
\begin{equation}
\label{eq:h1}
2 c_2 e^{-\rho } \rho +c_3 e^{\rho } \rho =(\rho^2 h_{(1)}')'
\end{equation}
which can in turn be integrated:
\begin{eqnarray}
%A_{(1)} &=& \frac{1}{r} \left(\int dr\; r  Y_{(1)}\right)\cr
%&=& \frac{c_1}{m^2} \frac{e^{m r}(m r-1)}{r} + \frac{c_2}{m^2} \frac{e^{-m r}(-m r-1)}{r}+ \frac{c_3}{r} \label{eq:sol1} \\
h_{(1)} &=&\frac{2 c_2 e^{-\rho }}{\rho }+\frac{c_3 e^{\rho }}{\rho }-\frac{c_4}{\rho }+c_5 \label{eq:h1sol}
\end{eqnarray}
 As discussed in \cite{Lu:2015psa}, we expect our solution to have 4 free parameters and indeed, substituting \eqref{eq:A1sol} and \eqref{eq:h1sol} back into the equations of motions shows that we must have $c_1=c_4$. Furthermore, requiring that our solution asymptote to flat space means one must eliminate the growing Yukawa mode and we  set $c_3=0$. The constant $c_5$ corresponds to a trivial rescaling of the time coordinate and requiring
  $\lim_{r \rightarrow \infty} h(r) =1$ sets $c_5=0$\footnote{The  constant, $c_5$, essentially maps to the constant, $c$, of the near horizon expansion in some non-trivial way.}. So finally the first order solution for the metric components is
\begin{eqnarray}
A&=&1 +\epsilon \left( \frac{c_1}{\rho }-\frac{c_2 e^{-\rho } (\rho +1)}{\rho }\right)+\ldots\label{eq:A1s}\\
h&=& 1+\epsilon  \left(  -\frac{c_1}{\rho} +2c_2 \frac{e^{-\rho}}{\rho}  \right) +\ldots  \label{eq:h1s} 
\end{eqnarray}
which is a special case of  results in \cite{stelle1978classical}. Given the boundary conditions we have imposed, at first order, there are two independent parameters in this expansion, $c_1$ and $c_2$.
The  parameter $c_1$ is related to the mass, while $c_2$ is related to the strength of the Yukawa mode
${e^{-m_2 r}}/{r}$. Recalling that, $\rho=m_2 r=r/L_\alpha$,  and letting 
\begin{eqnarray}
\label{eq:paramdef}
\epsilon L_\alpha c_1 &=& 2M\cr
\epsilon L_\alpha c_2 &=& \lambda
\end{eqnarray}
we see that we can consider an expansion in $M$ and $\lambda$ of the form 
\begin{eqnarray}
\label{eq:expansion2a}
h(r)&=& 1 + \sum_{
	\substack{n=1,2,\ldots\\0\leq i\leq n}
	} 
	\left(\frac{M}{L_\alpha}\right)^i\left(\frac{\lambda}{L_\alpha}\right)^{n-i} h_{(n,i)}(\rho)~, \cr
A(r)&=& 1 + \sum_{
		\substack{n=1,2,\ldots\\0\leq i\leq n}
	} 
	\left(\frac{M}{L_\alpha}\right)^i\left(\frac{\lambda}{L_\alpha}\right)^{n-i}A_{(n,i)}(\rho)~, \label{eq:expansion2b}
\end{eqnarray}
with $h_{(n,i)}(\rho)$ and $A_{(n,i)}(\rho)$ dimensionless.

To get a feel for the parameter $\lambda$, we note that the effective Newtonian potential seen by a particle far away from the black hole is
\begin{equation}
\label{eq:Newton}
\phi_N=\tfrac{1}{2}|g_{00}|\approx 1-\frac{M}{r}+\frac{\lambda e^{-r/L_\alpha}}{r}~.
\end{equation}

In practice, since the equations are linear at each order of $\epsilon$, one can also use the expansion \eqref{eq:expansion} and
easily read off powers of $M$ and $\lambda$ from \eqref{eq:paramdef}.  Notice that near the horizon (at $r=r_0$) one might roughly expect 
\begin{equation}
-\frac{\epsilon c_1}{r} = \frac{2M}{r_0} \sim 1~,\label{eq:c1est}
\end{equation}
even for small black holes. Similarly, for the other part of $g_{00}$, 
\begin{equation}
-\frac{\epsilon c_2  e^{-r/L\alpha} }{r} = \frac{2\lambda  e^{-r_0/L_\alpha} }{r_0}\sim ~ \frac{\lambda  e^{-2M/ L_\alpha} }{M},\label{eq:c2est}
\end{equation}
which is, a priori, difficult to estimate. 
Given that \eqref{eq:c1est} and  \eqref{eq:c2est} are not necessarily small near the horizon, one should not generically expect the expansion, \eqref{eq:expansion}, to be valid there. On the other hand, if $h_{(n,i)}(\rho)$ and $A_{(n,i)}(\rho)$ fall off fast enough, the expansion should become  good some distance from the horizon as these terms become small. In fact, we will find that as $|\lambda|$ increases, our expansion becomes worse and worse very near the horizon but rapidly converges as one moves just a small distance away from the horizon (measured in units of $L_\alpha$). 

Finally, we have a technical aside.
We find it convenient to introduce a third parameter in the asymptotic expansion, $h_\infty$, so that 
\begin{eqnarray}
	\label{eq:expansion2c}
	h(r)&=& h_\infty \left(1 + \sum_{
		\substack{n=1,2,\ldots\\0\leq i\leq n}
	} 
	\left(\frac{M}{L_\alpha}\right)^i\left(\frac{\lambda}{L_\alpha}\right)^{n-i} h_{(n,i)}(\rho)\right)~.
\end{eqnarray}
As discussed earlier, one is free to scale $h$ as long as one scales time to compensate. We would like to work in units for $t$ which give $h_\infty=1$, but one does not initially know how the near horizon expression  for $h$,  \eqref{anz2}, should be scaled to achieve this. For convenience, when setting the boundary conditions for our numerical integration, we fix $c=1$ in \eqref{anz2}. Fitting a value for $h_\infty$  to our numerical results then allows us to determine how we should rescale $t$ to ensure that $g_{00}|_{\rho\rightarrow\infty}=-1$. Scaling $t\rightarrow t/\sqrt{h_\infty}$, and using $c=1$, gives the modified form of \eqref{temp} we will be using
\begin{equation}
	\label{temp2}
	T= \frac{1}{2\pi L_\alpha }\sqrt{\frac{f_1}{h_\infty}}~.
\end{equation} 
\subsection{Second order expansion}

At second order, it is convenient to define the variables analogous to \eqref{eq:subs1}-\eqref{eq:subs2}:
\begin{eqnarray}
Y_{(2)} &=& \frac{1}{\rho}(\rho A_{(2)})' \label{eq:subs12}~,\\
Z_{(2)} &=& (\rho^2 h_{(2)}')'\label{eq:subs22}~.
\end{eqnarray}
Now, expanding the equations of motion \eqref{eq:EOM1} to second order we find
\begin{equation}
Z_{(2)}=2 \rho Y_{(2)}+ F_{(2)}~,\label{eq:EOM32}
\end{equation}
where
\begin{eqnarray}
F_{(2)} &=& c_2^2 e^{-2\rho}\left(\rho +8  +\tfrac{6  }{\rho}+\tfrac{3 }{\rho^2}\right)
-c_1 c_2  e^{-\rho}\left(\tfrac{7}{2} +\tfrac{9  }{2 \rho}+\tfrac{9}{2 \rho^2}\right)
   +c_	1^2\left(\tfrac{2 }{\rho^2}\right)~.
   \label{F2}
\end{eqnarray}
Following the same pattern as before, \eqref{eq:EOM32} can be substituted into \eqref{eq:EOM2} (together with \eqref{eq:subs12}, \eqref{eq:subs22}) to give (at second order)
\begin{equation}
\label{eq:y2}
 Y_{(2)}-Y_{(2)}''=\mathcal{D}A_{(2)}=J_{(2)}~,
\end{equation}
where
\begin{eqnarray}
\label{eq:J}
J_{(2)}(\rho)&=&      c_2^2 e^{-2 \rho}\left(\tfrac{27 }{4 \rho^5}+\tfrac{27 }{2 
   \rho^4}+\tfrac{71}{4 \rho^3}+\tfrac{35  }{2 \rho^2}+\tfrac{51 }{4 \rho}+\tfrac{15}{2}\right)\cr
   &&-c_1 c_2 e^{-\rho }\left(\tfrac{21}{
   \rho^5}+\tfrac{21 }{ \rho^4}+\tfrac{12 }{\rho^3}+\tfrac{5  }{\rho^2}+\tfrac{1}{\rho}\right)
   +c_1^2 \left(\tfrac{12}{ \rho^5}-\tfrac{1}{\rho^3}\right)~,
\end{eqnarray}
and once again $\mathcal{D}$ is given by \eqref{eq:A1b}.
Notice that \eqref{eq:y2} has the same form as \eqref{eq:A1} except that there is now  a source. We can solve \eqref{eq:y2} by  integrating the source over an appropriate  Green's function,  leading to
\begin{eqnarray}
\label{eq:A2sol}
A_{(2)} &=& \mathcal{D}^{-1} J_{(2)} \cr
 &=& 
 \tfrac{e^{-\rho } (\rho +1)  }{2 \rho }
 \int^\rho  ds\, e^{s } J_{\text{(2)}}(s ) 
 -\tfrac{e^{\rho } (1-\rho )  }{2 \rho }
 \int^\rho  ds\, e^{-s } J_{\text{(2)}}(s ) 
 -\tfrac{1}{\rho }
 \int^\rho ds\, s    J_{\text{(2)}}(s )  
  \cr
 &&+ \tfrac{ k_1 \, e^{-\rho } (\rho +1)  }{2 \rho }
    -\tfrac{k_2 \, e^{\rho } (1-\rho )  }{2 \rho } 
   - \tfrac{k_3 }{\rho } 
\end{eqnarray}
%The first two terms are the homogeneous solutions to \eqref{eq:y2} (i.e. in the absence of a source) corresponding to the two Yukawa modes which already appeared at first order.  The This means that we should choose $a_\pm$ and $\rho_\pm$ so that $Y_{(2)}$ has no homogeneous part.
The terms involving $k_i$ are just solutions to the homogeneous equation that appeared at first order -- without loss of generality we can set them to zero. In a similar fashion,  we do not specify the lower limits of the integrations since any constant pieces can be absorbed into the homogeneous solution.
Performing the integrals (and discarding the homogeneous part) we find
\begin{eqnarray}
\label{eq:A2sol}
A_{(2)} &=& 
c_2^2 e^{-2 \rho }  \left(\tfrac{69}{64}e^{3\rho } \text{Ei}(-3 \rho ) \left(\tfrac{1}{\rho }-1\right) -\tfrac{69}{64} e^{\rho }\text{Ei}(-\rho ) \left(\tfrac{1}{ \rho }-1\right) + \tfrac{9}{16
   \rho ^2}+\tfrac{21}{16 \rho }+\tfrac{5}{4}\right) \qquad \cr
   &&+ c_1 c_2 e^{-\rho } \left(\tfrac{1}{2}e^{2\rho } \text{Ei}(-2 \rho ) \left(1-\tfrac{1}{ \rho }\right)-\tfrac{7}{4 \rho ^2}-\tfrac{3}{2
   \rho }+\tfrac{1}{2}\log \rho\left(1+\tfrac{1}{ \rho }	\right)  \right)
   +\frac{c_1^2}{\rho ^2}
\end{eqnarray}
where $\text{Ei}(-x)$ is the exponential integral defined by 
\begin{equation}
\label{def:Ei}
\text{Ei}(-x)=-\int^\infty_x du\;\frac{e^{-u}}{u}\sim  -\frac{e^{-x}}{x}~\text{(for large $x$)}~.
\end{equation} 
To get some intuition for the terms involving Ei appearing in \eqref{eq:A2sol}, note that for large $x$,
\begin{eqnarray}
\label{eq:Eiprop}
-e^x\text{Ei}(-x)&\sim& \tfrac{1}{x}+\mathcal{O}(x^{-2})~,
\end{eqnarray}
whereas, for small $x$,
\begin{equation}
\label{eq:Eiprop2}
-e^x\text{Ei}(-x)\sim \gamma+\log (x )+\mathcal{O}(x)~,
\end{equation}
where $\gamma$ is Euler's constant.
Now that we have $A_{(2)}$, using \eqref{eq:subs12}-\eqref{F2},  we can find  $h_{(2)}$:
\begin{eqnarray}
%A_{(2)}&=&\frac{1}{r}\int dr \; mr Y_{(2)}= \frac{1}{\rho}\int d\rho \; \rho Y_{(2)} \\
h_{(2)}&=&\int d\rho \left(\frac{\int d\rho \ Z_{(2)}}{\rho^2} \right) 
%=\int d\rho \left(\frac{\int d\rho\; (2\rho Y_{(2)}+m^2F _{(2)})}{\rho^2} \right) ~.
=\int d\rho \left(\frac{2\rho A_{(2)} +\int d\rho\; F _{(2)}}{\rho^2} \right) \cr
&=&  c_2^2 e^{-2 \rho }\left(-\frac{69 e^{3\rho } \text{Ei}(-3 \rho )}{32 \rho }+\frac{3  e^{2 \rho }\text{Ei}(-2 \rho )}{2}+\frac{69 e^{\rho } \text{Ei}(-\rho )}{32 \rho }+\frac{15}{16 \rho ^2}-\frac{1}{4\rho }\right) 
   \cr 
   &&+c_1 c_2 \left(\frac{e^{\rho } \text{Ei}(-2 \rho )}{\rho }+e^{-\rho } \left(-\frac{1}{2 \rho ^2}-\frac{\log (\rho )}{\rho }\right)\right)
\label{eq:h2sol}
\end{eqnarray}
The integrations in \eqref{eq:h2sol}  produce two constants of integration which we neglected. One of the constants can be absorbed into the mass and the other corresponds to a rescaling of the time coordinate so that no new degrees of freedom are introduced at this order. For more details of the calculation see  \autoref{ap:asymp}.

\subsection{Higher order expansion}
At higher orders, we can make the substitutions
\begin{eqnarray}
Y_{(n)} &=& \frac{1}{\rho}(\rho A_{(n)})'~, \label{eq:subs1n}\\
Z_{(n)} &=& (\rho^2 h_{(n)}')'\label{eq:subs2n}~.
\end{eqnarray}
and it would appear that the equations to order $n$,  take the form
\begin{eqnarray}
Z_{(n)}&=&2 \rho Y_{(n)}+ F_{(n)}~,\label{eq:EOM3n} \\
 Y_{(n)}-Y_{(n)}''&=&\mathcal{D} A_{(n)}= J_{(n)}(\rho)\label{eq:yn}~,
\end{eqnarray}
for some $F_{(n)}$ and $J_{(n)}$. As at second order, we find a solution of the form
\begin{eqnarray}
\label{eq:ynsol1}
% m^{-n}Y_{(n)} &=& \tfrac{1}{2}e^{-\rho}\int^\rho  dk\; e^{k}J_{(n)}(k) - \tfrac{1}{2} e^{\rho}\int^\rho   dk\; e^{-k}J_{(n)}(k))~,\cr
A_{(n)}&=& %\frac{1}{\rho}\int d\rho \; \rho Y_{(n)}~,
 \tfrac{e^{-\rho } (\rho +1)  }{2 \rho }
 \int^\rho  ds\, e^{s } J_{{(n)}}(s ) 
 -\tfrac{e^{\rho } (1-\rho )  }{2 \rho }
 \int^\rho  ds\, e^{-s } J_{{(n)}}(s ) 
\cr 
&&-\tfrac{1}{\rho }
 \int^\rho ds\, s    J_{{(n)}}(s )  
 \\
h_{(n)}&=& 
\int d\rho\left(\frac{2\rho A_{(n)} +\int d\rho\; F _{(n)}}{\rho^2} \right) ~.\label{eq:ynsol2}
\end{eqnarray}
We note that, as for the second order case, no new degrees of freedom are introduced at each order.
Unfortunately, in practice, the integrals become increasingly difficult to
evaluate due to non-trivial integrals involving Ei and Meijer-G functions, although
in principle, it seems one could evaluate up to arbitrary order using numerical integration.
More details, including the $3^\text{rd}$ order solutions, are in  \autoref{ap:asymp}. 

\subsection{Features of the asymptotic expansion}\label{sec:feat}

We would like to investigate some features of the asymptotic expansion.
When all the dust has settled, we find that the expression for the metric components to second order in $M$ and $\lambda$ is
\begin{eqnarray}
h&=& 1-\frac{2  M}{r } +\frac{2 \lambda  e^{-r/L_\alpha }}{r }
+\left(\frac{ M\lambda}{L_\alpha^2}\right) e^{-\rho }
\left(
\tfrac{2 e^{2\rho } \text{Ei}(-2 \rho )}{\rho }-\tfrac{2  \log \left({\rho }\right)}{\rho}-\tfrac{1}{\rho ^2}
\right)\cr
&&+\left(\frac{ \lambda^2}{L_\alpha^2}\right) e^{-2 \rho } 
\left(
\tfrac{69 e^{\rho }\text{Ei}(-\rho )}{32 \rho }
+\tfrac{3 e^{2 \rho }\text{Ei}(-2 \rho )}{2}
-\tfrac{69 e^{3\rho } \text{Ei}(-3 \rho )}{32 \rho }
+\tfrac{15 }{16 \rho ^2}
-\tfrac{1}{4 \rho }
\right)+\ldots\label{eq:O2a}\\
A&=&1+\frac{2  M}{r }+\frac{4M^2}{r^2}- \left(\frac{\lambda}{L_\alpha}\right) e^{-\rho}\left(1+\tfrac{1}{\rho }\right)\cr
&&+   \left(\frac{ M\lambda}{L_\alpha^2}\right) e^{-\rho }
\left(
e^{2\rho }\text{Ei}(-2 \rho )\left(1-\tfrac{1}{\rho }\right)
+ \log {\rho }\left(1+\tfrac{1}{\rho }\right)
-\tfrac{7 }{2 \rho^2}
-\tfrac{3 }{\rho }
\right)\cr
&&+ \left(\frac{\lambda^2}{L_\alpha^2}\right) e^{-2 \rho }
\left(
\tfrac{5}{4}-\tfrac{69}{64}e^{3\rho } \text{Ei}(-3 \rho )\left(1-\tfrac{1}{\rho }\right) 
-\tfrac{69 }{64}e^{\rho }\text{Ei}(-\rho )\left(1+\tfrac{1}{\rho }\right)
+\tfrac{9 }{16 \rho ^2}
+\tfrac{21}{16 \rho }\right)\cr
&&+\ldots\hfill\label{eq:O2b}
\end{eqnarray}
Notice that with $\lambda=0$, it is easy to check that we recover the Schwartzchild solution (up to $\mathcal{O}(M^2)$):
%%%%%%%%%%
\begin{eqnarray}
\label{eq:Sw_exp1}
h_\text{Schwarzschild }&=&1-\frac{2  M}{r } \\
A_\text{Schwarzschild }&=&\left(1-\frac{2  M}{r }\right)^{-1}=1+\frac{2  M}{r }+\frac{4  M^2}{r^2}+\frac{8M^3}{r^2 }+\ldots \label{eq:Sw_exp2}
\end{eqnarray}
%%%%%%%%%
At large $r$, we find the slightly simpler expressions
\begin{eqnarray}
h&=& 1-\frac{2  M}{r } +\frac{2 \lambda  e^{-r/L_\alpha }}{r }
+ \left(\frac{ M\lambda}{L_\alpha^2}\right)e^{-\rho }
\left(
-\tfrac{2 \log \left({\rho }\right)}{\rho }-\tfrac{2}{\rho ^2}+\mathcal{O}({\rho^{-3} })
\right)\cr
&&+  \left(\frac{ \lambda^2}{L_\alpha^2}\right) e^{-2 \rho } 
\left(
-\tfrac{1}{\rho }-\tfrac{1}{8 \rho ^2}+\mathcal{O}({\rho^{-3} })
\right)+\ldots~,\label{eq:larger1}\\
A&=&1+\frac{2  M}{r }+\frac{4M^2}{r^2}-\left(\frac{ \lambda}{L_\alpha}\right) e^{-\rho}\left(1+\tfrac{1}{\rho }\right)\cr
&&+   \left(\frac{ M\lambda}{L_\alpha^2}\right) e^{-\rho }
\left(
\log (\rho )\left(1+\tfrac{1}{\rho }\right)
-\tfrac{7}{2 \rho }
-\tfrac{11}{4 \rho ^2} +\mathcal{O}({\rho^{-3} })
\right)\cr
&&+ \left(\frac{ \lambda^2}{L_\alpha^2}\right)  e^{-2 \rho }
\left(\tfrac{5}{4}+\tfrac{11}{4 \rho }+\tfrac{1}{12\rho ^2}+\mathcal{O}({\rho^{-3} })
\right)\cr
&&+\ldots~.\hfill\label{eq:larger2}
\end{eqnarray}
We note that the expansions \eqref{eq:larger1} and \eqref{eq:larger2} look like transseries (see \cite{trans} for a nice introduction to transseries).  
Perhaps starting with an appropriate transseries ansatz would be an easier way to approach this problem which we hope to look at in the future. Given the appearance of Yukawa modes  at first order, it is perhaps not surprising that we end up with a transseries expansion \& and we conjecture that this feature will be a common characteristic of the asymptotics of other higher derivative black holes.

\begin{figure}[htb]
	\includegraphics[width=0.95\linewidth]{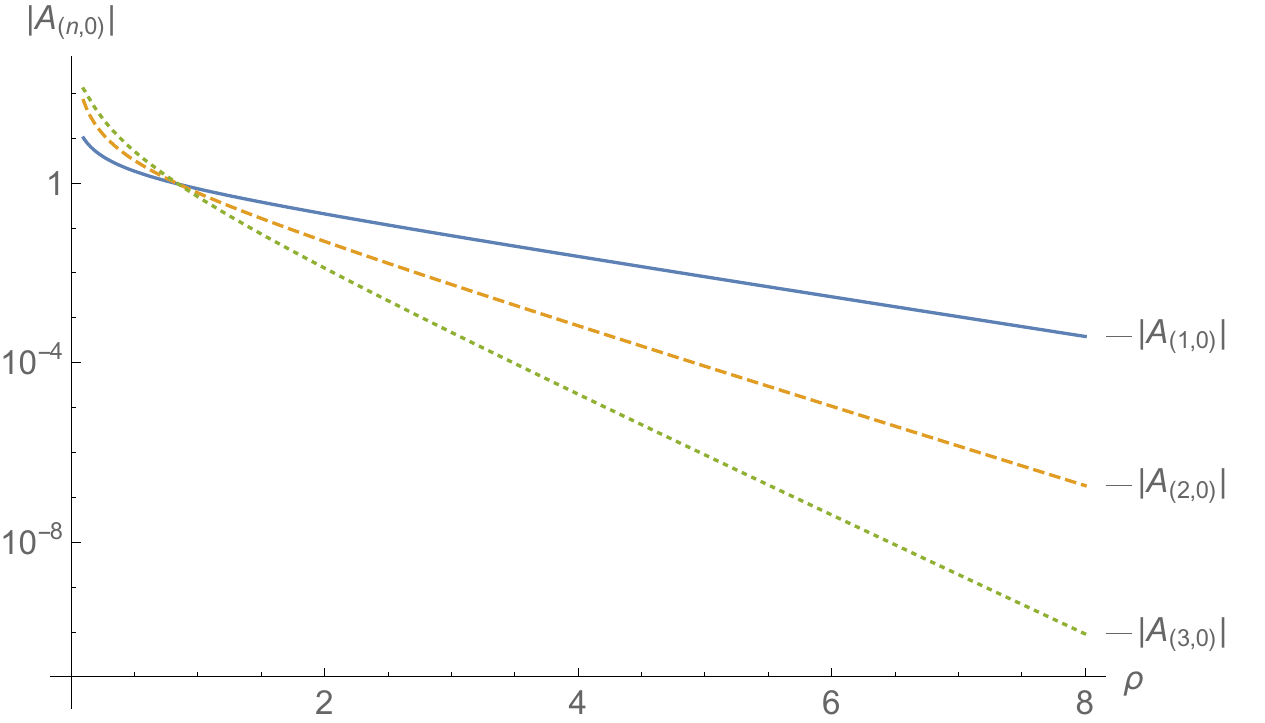}
	\caption{Log plot of $|A_{(n,0)}|$ for $n=1,2,3$. In the expansion, \eqref{eq:expansion2b},  $A_{(n,0)}$ is multiplied by a factor $M^0 (\lambda/L_\alpha)^{n}$, i.e.  $|A_{(n,0)}|$  are the pieces of the metric which depend only on $\lambda$ and not $M$. The solid line is $|A_{(1,0)}|$, the dashed line is $|A_{(2,0)}|$ and the dotted line is $|A_{(3,0)}|$.}
	\label{fig:An0}
\end{figure}
\begin{figure}[htb]
	\includegraphics[width=0.95\linewidth]{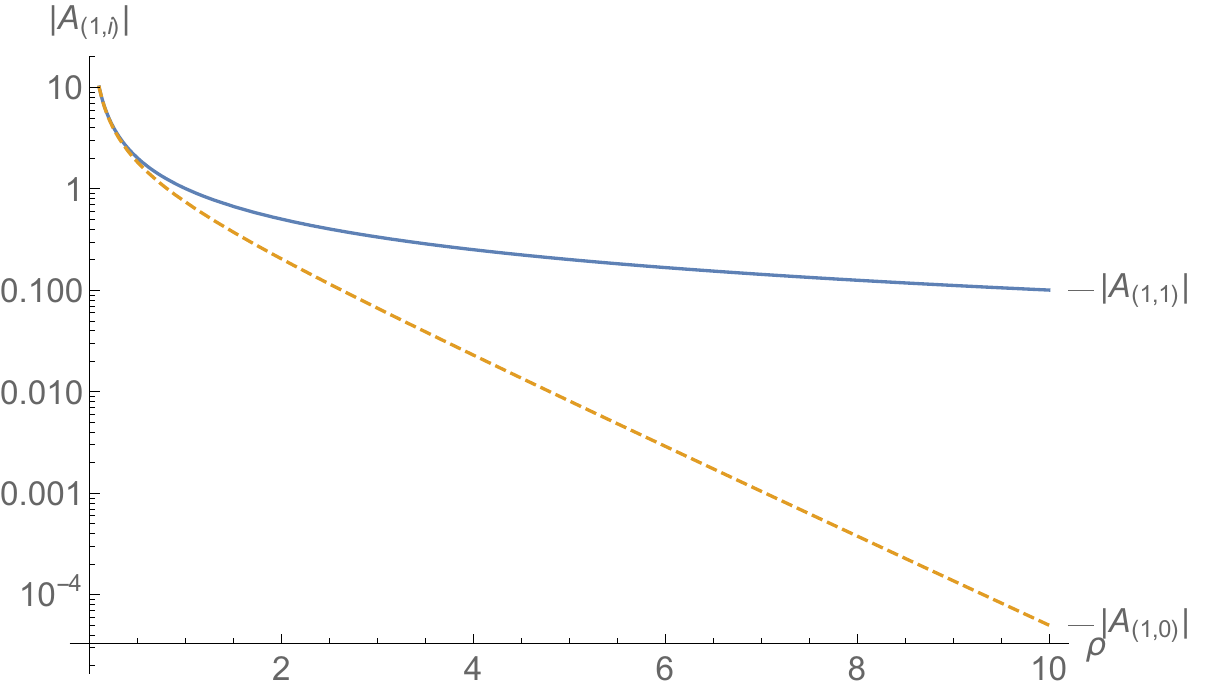}
	\caption{Log plot of the first order terms $|A_{(1,i)}|$ for $i=0,1$. In the expansion, \eqref{eq:expansion2b},  $A_{(1,i)}$ is multiplied by a factor $M^i \lambda^{(1-i)}$.  The solid line is $|A_{(1,1)}|$   and the dotted line is $|A_{(1,0)}|$.
		For the Schwarzschild black hole, one only has the term $A_{(1,1)}$ at this order.
	}
	\label{fig:A1i}
\end{figure}
\begin{figure}[htb]
	\includegraphics[width=0.95\linewidth]{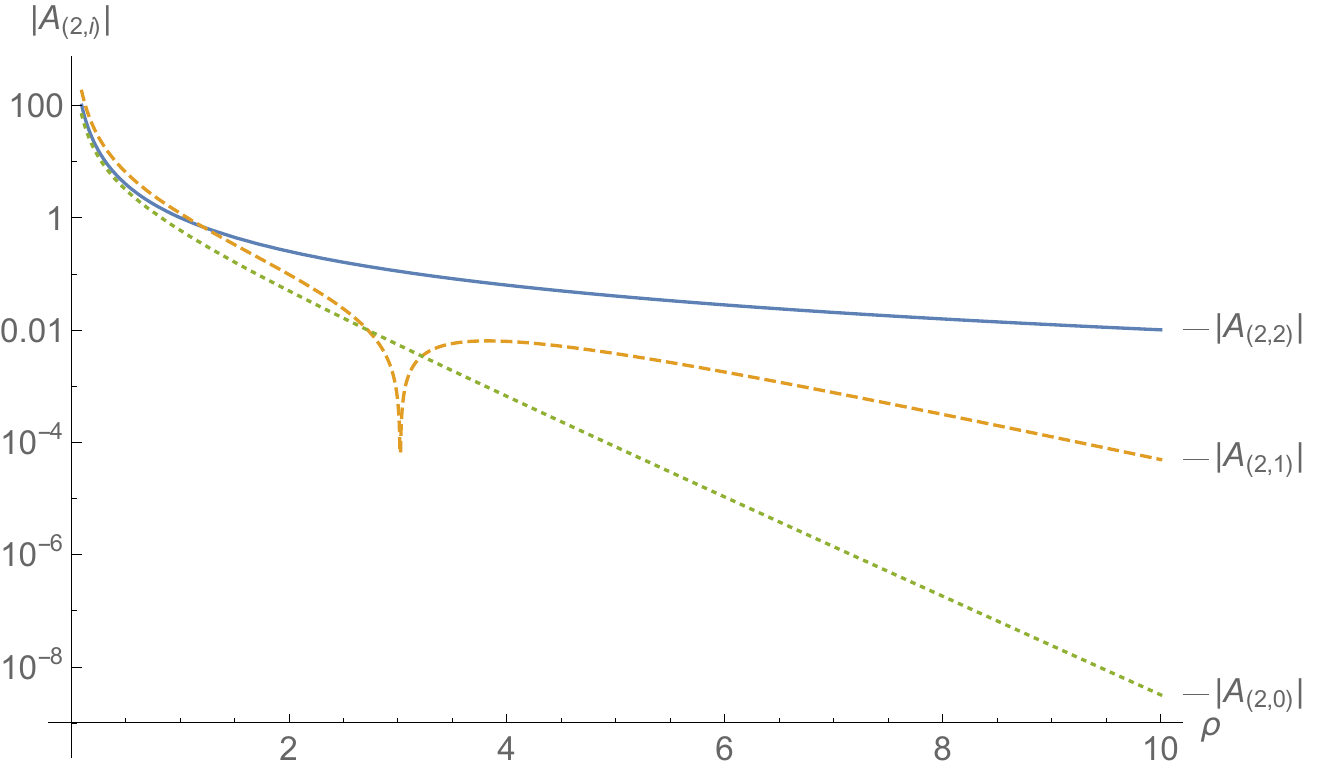}
	\caption{Log plot of the second order terms $|A_{(2,i)}|$ for $i=0,1,2$. In the expansion, \eqref{eq:expansion2b},  $A_{(2,i)}$ is multiplied by a factor $M^i \lambda^{(2-i)}$.  The solid line is $|A_{(2,2)}|$, the dashed line is $|A_{(2,1)}|$  and the dotted line is $|A_{(2,0)}|$.
		For the Schwarzschild black hole, one only has the term, $A_{(2,2)}=\frac{4}{\rho^2}$, at this order. The funny blip in $|A_{(2,1)}|$ corresponds to a sign change in $A_{(2,1)}$.
	}
	\label{fig:A2i}
\end{figure}
\begin{figure}[htb]
	\includegraphics[width=0.95\linewidth]{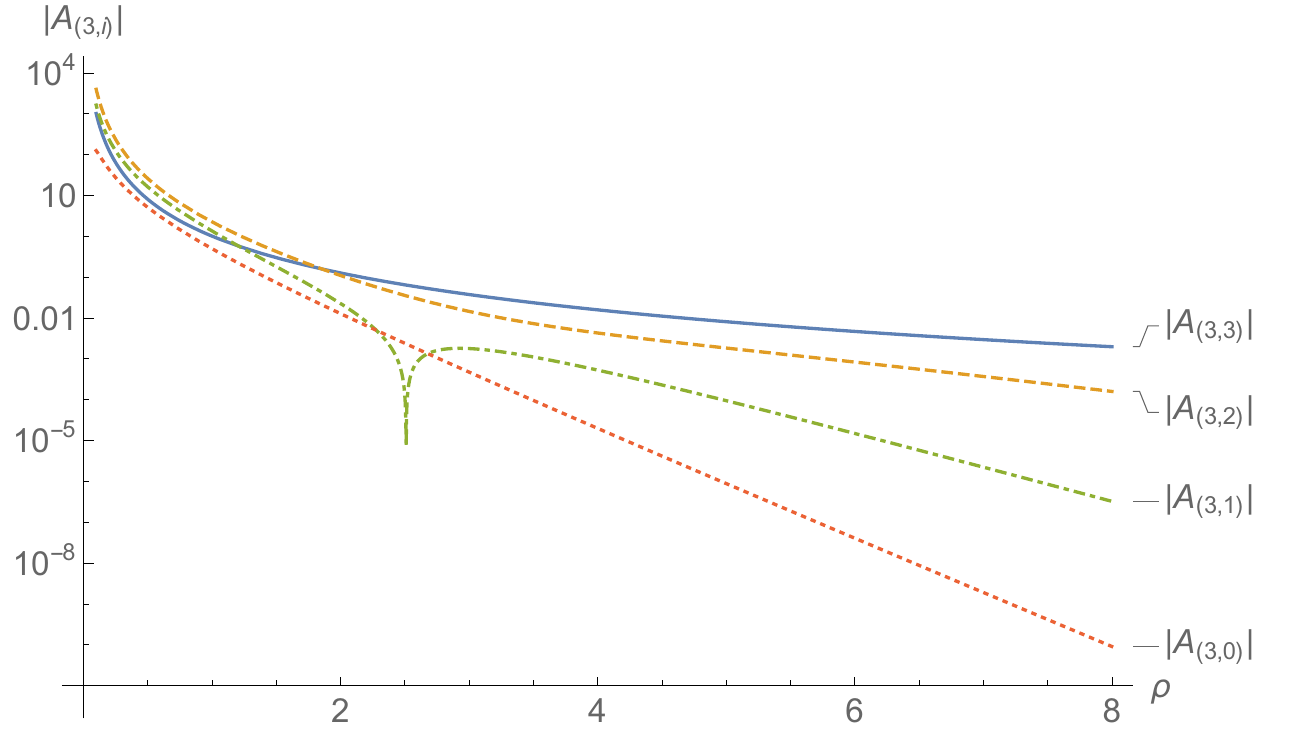}
	\caption{Log plot of the third order terms $|A_{(3,i)}|$ for $i=0,1,2,3$. In the expansion, \eqref{eq:expansion2b},  $A_{(3,i)}$ is multiplied by a factor $M^i \lambda^{(3-i)}$.  The solid line is $|A_{(3,3)}|$, the dashed line is $|A_{(3,2)}|$, the dash-dotted line is $|A_{(3,1)}|$ and the dotted line is $|A_{(3,0)}|$.
		For the Schwarzschild black hole, one only has the term $A_{(3,3)}=\frac{8}{\rho^3}$ at this order. The funny blip in $|A_{(3,1)}|$ corresponds to a sign change in $A_{(3,1)}$.
	}
	\label{fig:A3i}
\end{figure}

To get a feel for expansion we have numerically plotted various terms in  the expansion  of $A$,  \eqref{eq:expansion2a}, in \autoref{fig:An0}-\autoref{fig:A3i}.  
There are 3 notable features of these plots we would like to emphasise 
\begin{enumerate}
	\item \autoref{fig:An0} is a log plot of  $A_{(1,0)}$, $A_{(2,0)}$, and $A_{(3,0)}$ as a function of $\rho=r/L_\alpha$. These are terms in \eqref{eq:expansion2b}  with coefficients $\lambda$, $\lambda^2$ and $\lambda^3$ that do not involve the mass $M$. It is clear from the plot that as $\rho$ increases, successive terms become exponentially smaller at higher orders in the expansion. There is similar behaviour when one compares other higher order terms in the expansion to the lower order term which effectively act as sources for them.  This is good evidence that the expansion will converge at least  for sufficiently large $\rho$ -- we leave a more rigorous examination of convergence for possible future work. The plots in \autoref{fig:An0} all approach straight lines indicating exponential falloff on a log plot.    
	\item \autoref{fig:A1i}-\autoref{fig:A3i} are plots of different terms at the same order in the asymptotic expansion. What is clear is that for $\rho>>1$, the terms $A_{(n,n)}$, which correspond to the pure Schwarzschild black hole, dominate for large $\rho$. Far from the horizon, the non-Schwarzschild terms become negligible. The non-Schwarzschild terms also become linear at large $\rho$ once again indicating exponential falloff.     
	\item From \autoref{fig:An0}, one see that for small $\rho$, higher order terms in the expansion are larger than the lower order terms. This suggests that the expansion breaks down at some point. We know that the expansion for the Schwarzschild black hole only breaks down at the horizon in these coordinates (which would be at $\rho=1$). We calculate the asymptotic expansion up to third order and our numerical investigations show that as $\lambda$ becomes large, the asymptotic solution does not match well with the near horizon expansion.  
	 Consequently, we will need numerical integration to match the near horizon solution with the asymptotic expansion which is what we cover in the next section. 
\end{enumerate}

\section{Matching  and Thermodynamics}\label{sec:match} 

 In this section we study the non-Schwarzschild thermodynamics, as well as the validity of our asymptotic expansion. At the most basic level, we need to find the relationship between the mass of the solution, which can be extracted asymptotically, and, temperature and entropy, which are encoded in the horizon data.

As discussed in section {\ref{sec:NH}}, the near horizon solution is characterised by two parameters $\delta$ and $\rho_0$ with the former encoding the deviation for the Schwarzschild solution and the latter encoding the position of the horizon. As shown in \cite{Lu:2015cqa}, for a given $\rho_0$, aside from the Schwarzschild solution, which corresponds to $\delta=0$, there is one additional asymptotically flat solution with  $\delta\neq0$. \footnote{There is a special point at $\rho_0\approx0.876$ where the Schwarzschild solution is the only solution. In \cite{Lu:2015cqa}, the solutions with $\rho_0>0.876$ were found.} As discussed below, for a given $\rho_0$, one has to tune $\delta$ to obtain a regular solution.  In section \ref{sec:asymp} we showed that the asymptotic solution is characterised by two parameters, the mass $M$, and strength of the Yukawa excitation, $\lambda$ (see \eqref{eq:Newton}). The Schwarzschild solution corresponds to $\lambda=0$. We will discuss below how for a given non-Schwarzschild solution, one can extract $M$ and $\lambda$ by matching the near horizon and asymptotic expansions numerically. 

We start with a similar procedure to the one outlined in \cite{Lu:2015cqa}. For a fixed $\rho_0$, one uses the near horizon expansion to set the boundary conditions for numerical integration of \eqref{eq:ETr} and \eqref{eq:Err}, starting a little bit away from the horizon.  One then has to find $\delta$ so that the positive Yukawa mode, which would cause the solution to blow up, is not excited.
\begin{figure}[htb]
	\includegraphics[width=1\linewidth]{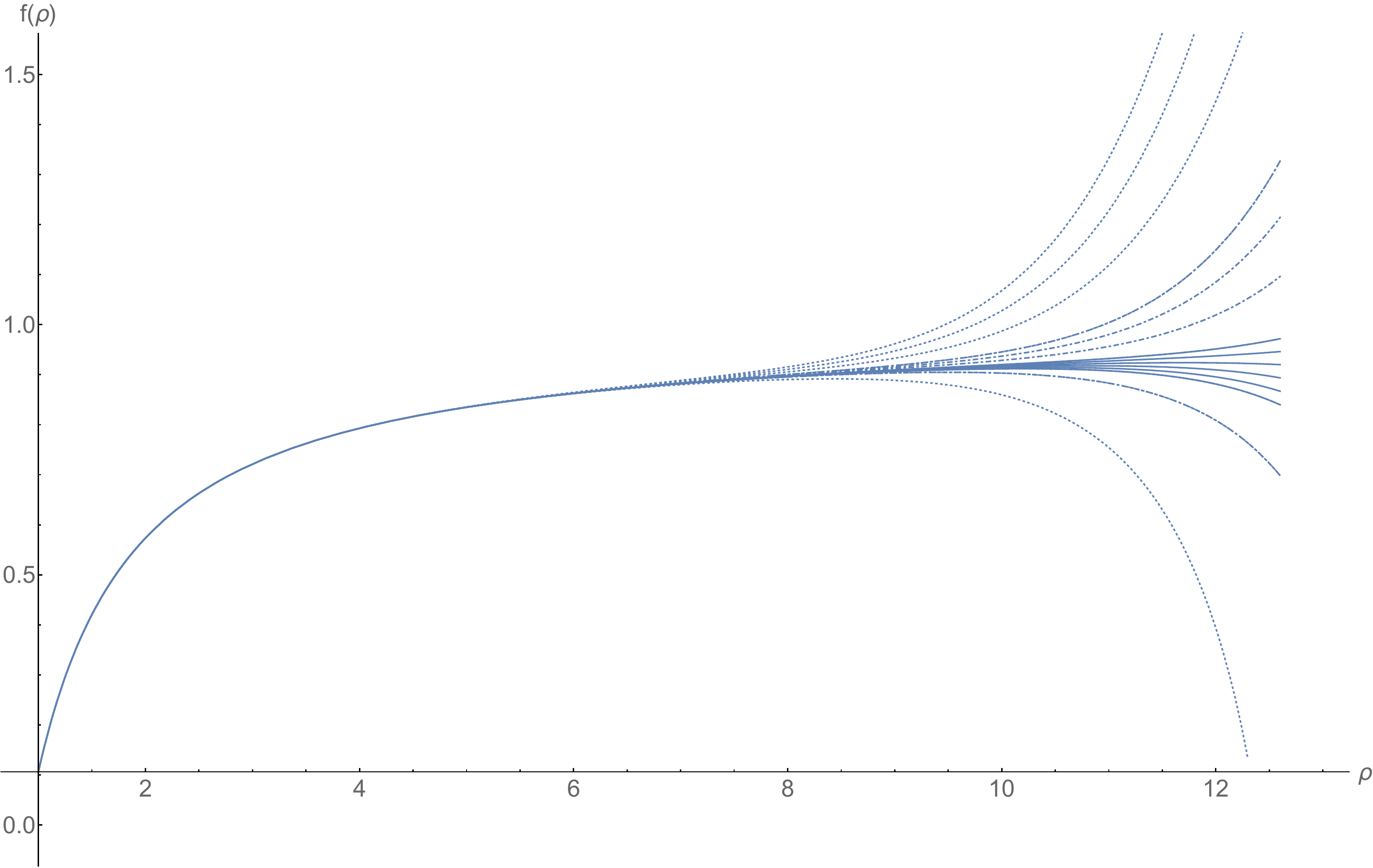}
	\caption{Numerical plot of $f(r)$ (with $\rho_0=0.9$) showing the procedure for tuning $\delta$ and extending how far one can integrate before the solution diverges. The dotted plots correspond to 
		$0.06584\leqslant\delta\leqslant 0.6616$. The dashed plots show the next step with $0.066032\leqslant\delta\leqslant 0.66096$ and the solid plots show a further step  with  $0.0660704\leqslant\delta\leqslant 0.0660832$.
	}
	\label{fig:fiter}
\end{figure}
Adopting a shooting method, we scan over a range of guesses for $\delta$, successively tuning it so that the integration can extend further and further -- this process is shown in \autoref{fig:fiter}.\footnote{We used Mathematica for the numerical integration with the stiffness switching method, successively increasing the precision and accuracy goals as we fine tuned $\delta$.}  It was claimed in \cite{Lu:2015cqa}, that they were able to  extend the numerics integration as far as $\rho=60$. While this  is good evidence for the existence of non-Schwarzschild solutions, there is no guarantee that one will eventually hit some sort of barrier.  We found that one does not have to integrate too far  before the asymptotic expansion, which also involves only three parameters, $M$, $\lambda$ and $h_\infty$, matches very well with  numerical integration\footnote{We used a non-linear fit to match the numerical results with the asymptotic expansion.}. This gives even stronger evidence for the non-Schwarzschild solutions. 

\subsection{Convergence and Asymptotics}

Before going on to consider the thermodynamics, we would like to discuss some investigations of the near horizon and asymptotic expansions, as well as our matching procedure. 

\begin{figure}[htb]
	\includegraphics[width=1\linewidth]{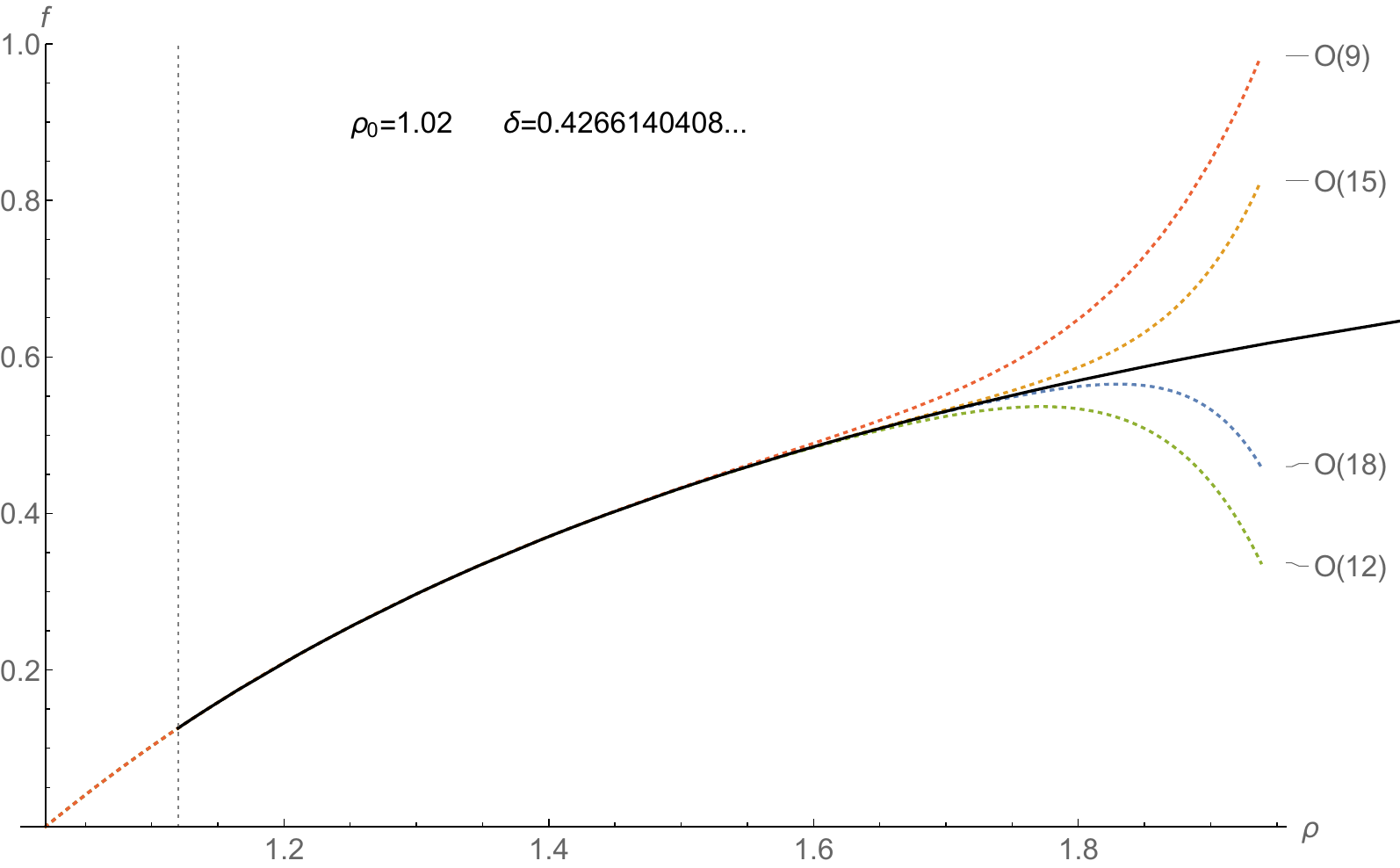}
	\caption{Plots of the near horizon expansion \eqref{anz3} for $f(\rho)$ compared with results of the numerical integration with  $\rho_0=1.02$ and  $\delta=0.4266\ldots$~. The numerical plot is shown as a solid line and the dashed lines are plots of the near horizon expansion up to orders 9,12,15 and 18 as labelled.  The $\mathcal{O}(18)$ expansion was used to define boundary conditions for the numerical integration which started  at $\rho=1.12$. The point at which we started the numerical integration is denoted by a dashed grey vertical line in the plot.   
	}
	\label{fig:pert1}
\end{figure}

In \autoref{fig:pert1} we show a numerical plot for $\rho_0=1.02$ and $\delta \approx 0.4266\ldots$, together with the near-horizon expansion at various orders. It is clear from the plot that the near horizon expansion breaks down pretty quickly, namely before $2\rho_0$,  even at high order. We see that one does not  greatly extend the expansion's domain of validity each time one increases the order of the expansion.  We found this behaviour was universal, over all values of $\rho_0$ investigated, with the near horizon expansion, even up to $\mathcal{O}(18)$, not extending far.

Given the computing power at our disposal, we did not extend our numerical integration much beyond 40-50 $\rho_0$. As mentioned, and as can be seen from \autoref{fig:An0} - \autoref{fig:A3i}, terms in our asymptotic expansion involving $\lambda$ fall of exponentially fast. This means that our solution is soon very well approximated by the Schwarzschild making it relatively easy to find the mass, $M$, and $h_\infty$. Using a non-linear fitting procedure we extract the parameters by finding  the best fit between our numerical and asymptotic solutions (over some appropriately chosen range of $\rho$). 

\begin{figure}[htb]
	\includegraphics[width=1\linewidth]{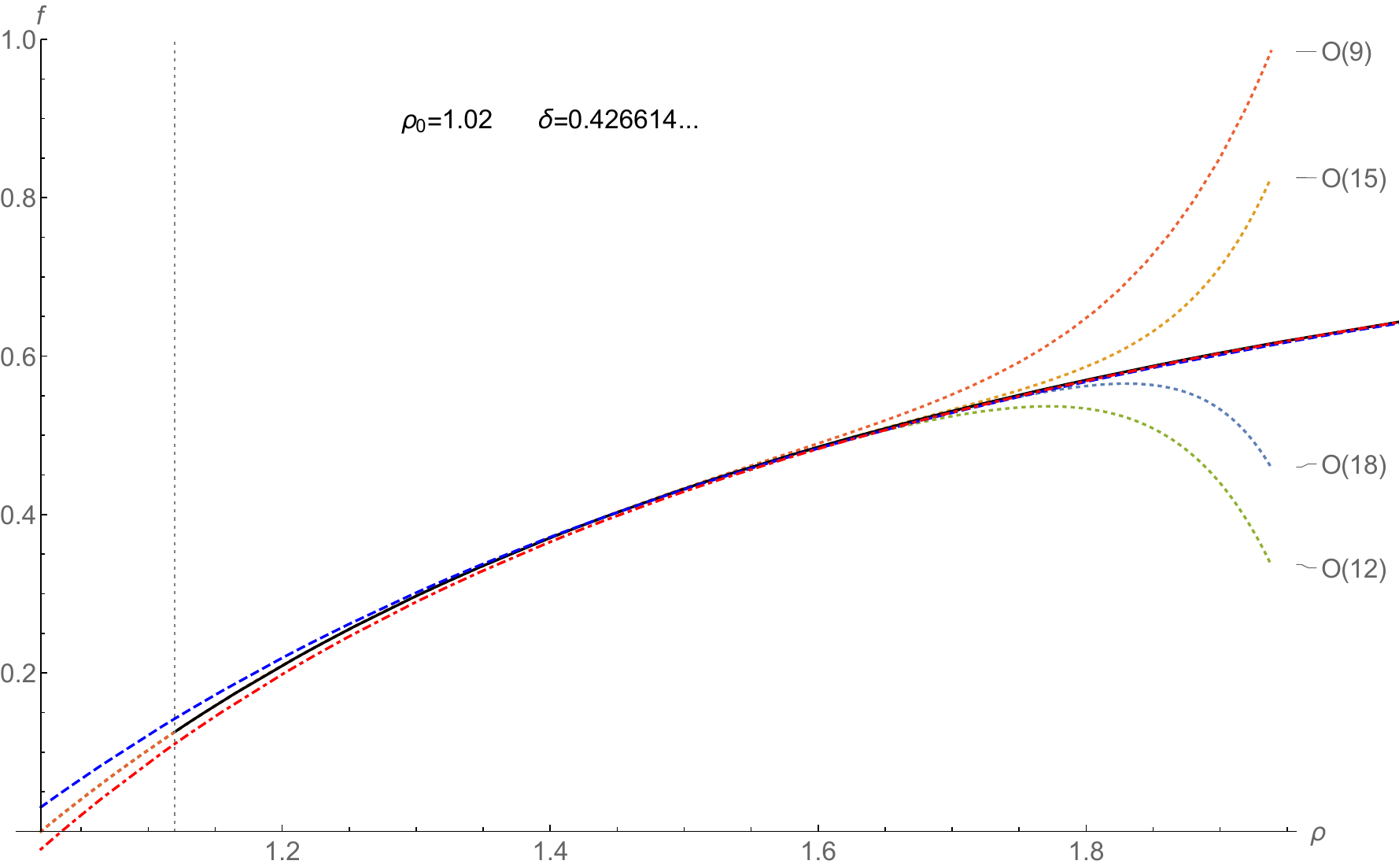}
	\caption{A repeat of \autoref{fig:pert1} with the fit of the $\mathcal{O}(2)$ (blue line with larger dashes) and $\mathcal{O}(3)$ (red dot-dashed line) asymptotic expansions added. 
	}
	\label{fig:pert1b}
\end{figure}

In  \autoref{fig:pert1b} we superimpose our fit of the $\mathcal{O}(2)$ and $\mathcal{O}(3)$ asymptotic expansions onto \autoref{fig:pert1} \footnote{The asymptotic expansions where fitted to the numerical results over the range $2.5\rho_0<\rho<28\rho_0$.}. We see that there is a good match even close to the horizon. Furthermore, although it is not too easy to see on this graph, the $\mathcal{O}(3)$ plot looks like a better fit. Unfortunately, we also found that  the asymptotic expansion is generally not  good in the limited domain of validity of the near horizon expansion. 
\begin{figure}[htb]
	\includegraphics[width=1\linewidth]{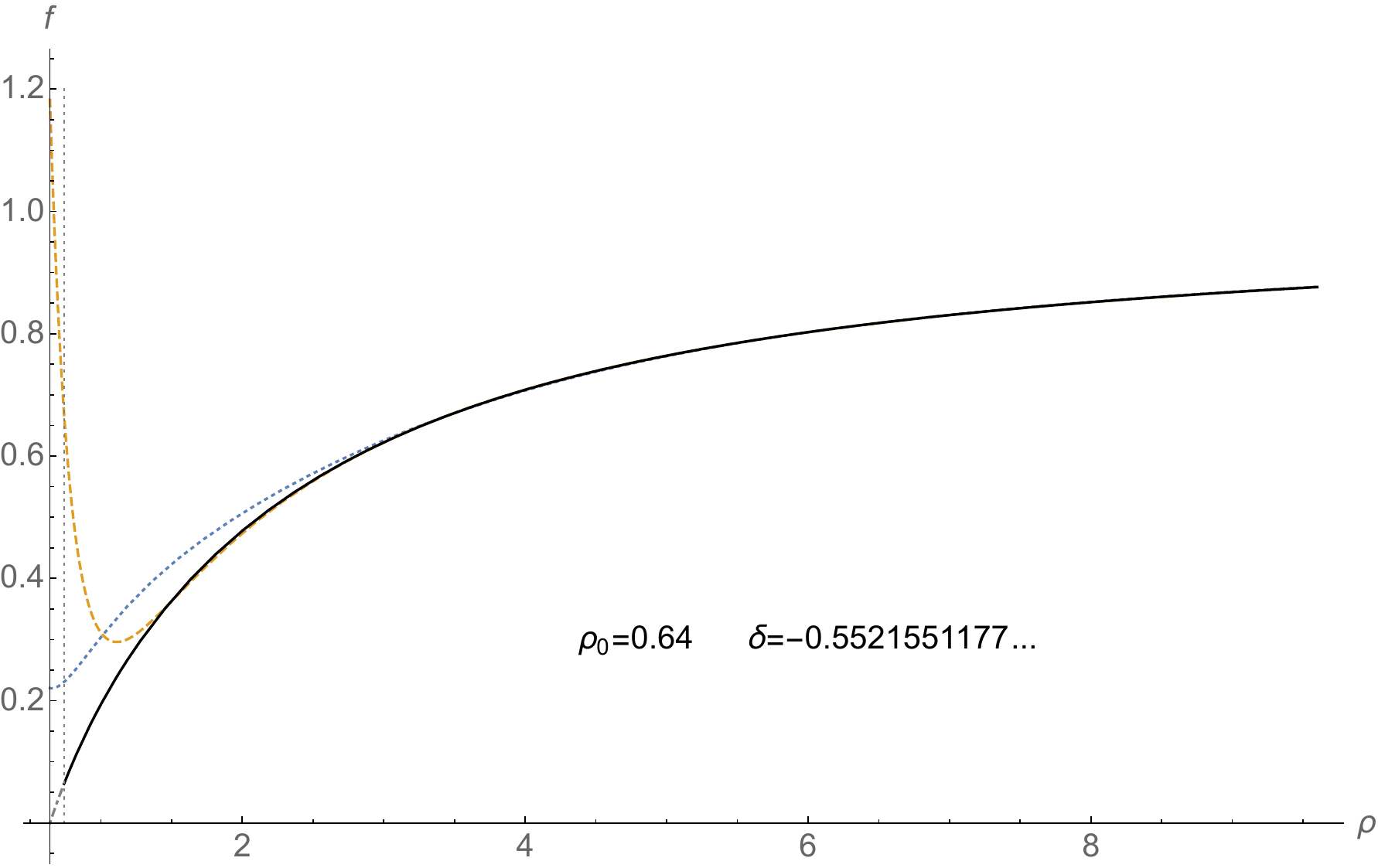}
	\caption{Plot of the numerical integration of $f(\rho)$ against fits of the second and third order asymptotic expansion. The numerical integration is denoted by a solid back line, the third order fit is a mustard dashed line and the second order fit is a blue dotted line. A vertical dotted line denotes where we started the numerical integration and the dash-dotted grey line is the near horizon expansion.  The horizon radius was $\rho_0=0.64$ and we found $\delta=-0.552155\ldots$. The fitted values for $\lambda$ at second and third order were,  
		 $\lambda_{\mathcal{O}(2)}=1.48\ldots$ and $\lambda_{\mathcal{O}(3)}=0.42\ldots$, respectively. The asymptotic expansions were fitted to the numerical results over the range $\rho_*=4 \rho_0$ to $\rho_f=40\rho_0$.
	}
	\label{fig:numvsasymp}
\end{figure} 
This is evident in \autoref{fig:numvsasymp} where we've plotted  numerical results for $\rho_0=0.64$ together with fits for the asymptotic expansions. 
We see that the $\mathcal{O}(3)$ expansion diverges more from the near horizon solutions than the $\mathcal{O}(2)$ expansion  but on the other hand the $\mathcal{O}(3)$ expansion converges more rapidly to the numerical results as $\rho$ increases. We find similar results for other values of $\rho_0$ with the divergence between the asymptotic and near horizon expansion increasing as $|\lambda|$ increases (where $\lambda$ is obtained from our best fit procedure). This behaviour is not too surprising if we look back at \Cref{fig:An0,fig:A1i,fig:A2i,fig:A3i}, where terms which appear with a coefficient of $\lambda$ can be very large for small $\rho$ but fall off very fast. Putting this all together strongly suggests that expansion \eqref{eq:expansion2b} becomes good for $\rho$ sufficiently large. On the other hand, this means that  in general, to find the relationship between near horizon and asymptotic data some numerical integration is required. Lastly, we would like to note that while the values obtained for $M$ and $h_\infty$ by fitting the $\mathcal{O}(2)$ or $\mathcal{O}(3)$ asymptotic expansions to the numerical results only differ by 1\% or less, the values obtained for $\lambda$ can have much larger variation. For example, for the case shown in \autoref{fig:numvsasymp}, fitting the second order expansion gave $\lambda_{\mathcal{O}(2)}\approx 1.48$ while fitting the third expansion gave   $\lambda_{\mathcal{O}(3)}\approx0.42$. Once again, we can understand this from the fact that terms involving $\lambda$ fall off very quickly leaving an essentially Schwarzschild like solution.   

While we can see from \Cref{fig:pert1b,fig:numvsasymp} that the numerical and asymptotic results agree well beyond a few multiples of $\rho_0$, to get a better feel for our results we now switch to log-log plots to make small differences at large $\rho$ more visible.    
\begin{figure}[htb]
	\includegraphics[width=1\linewidth]{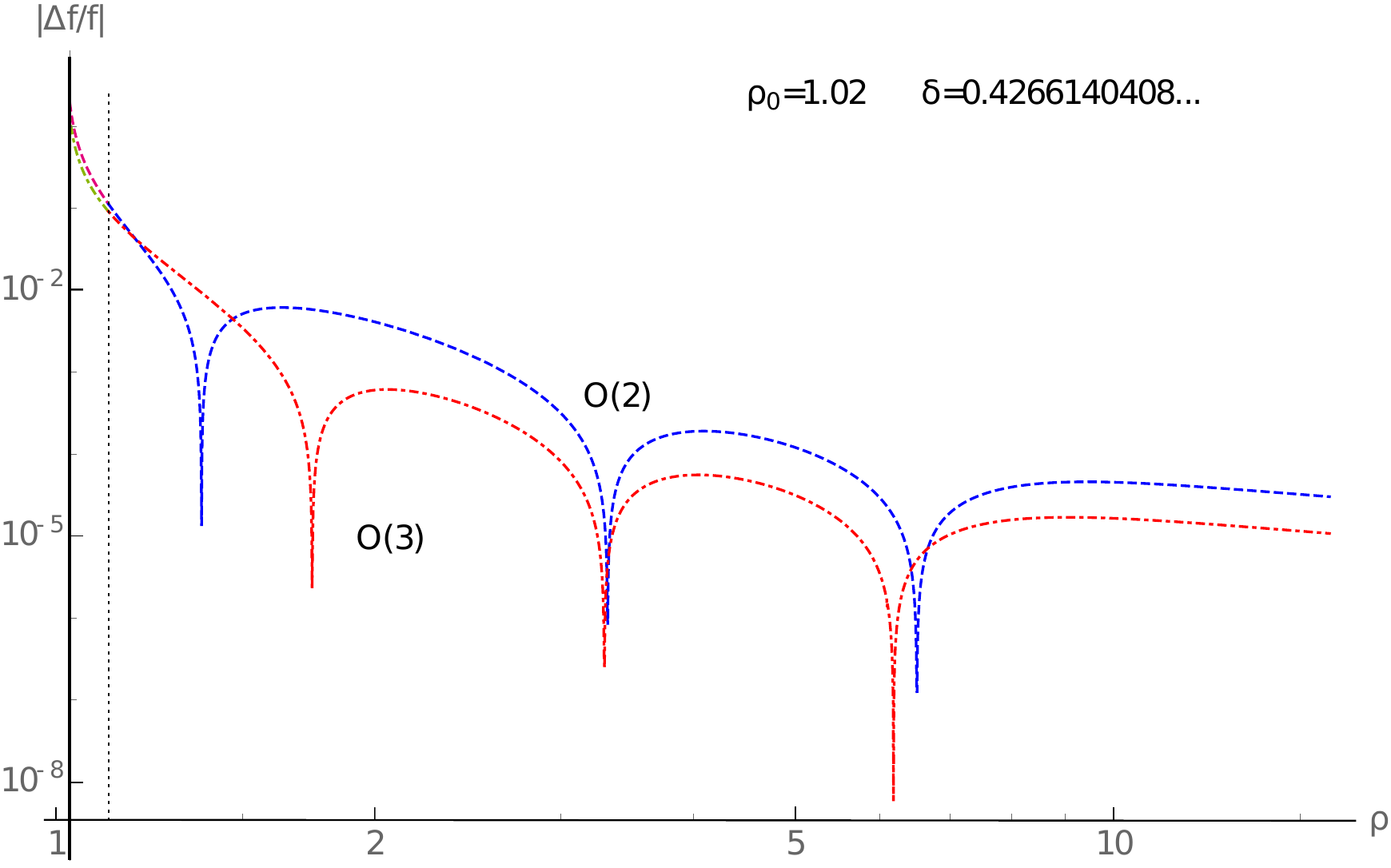}
	\caption{Log-log plots of ${|\Delta f|}/{f}$ against $\rho$ for the second (blue dashed line) and third order (red dot-dashed line) asymptotic expansions (see \eqref{plot def}-\eqref{fN_def} for the definition of $|\Delta f|/f$).
	The fits for $\lambda$ where, $\lambda_{\mathcal{O}(2)}=0.79\ldots$ and  $\lambda_{\mathcal{O}(3)}=0.71\ldots$, over the range $\rho_*=2.5\rho_0$ to $\rho_f=40\rho_0$.
	}
	\label{fig:log1}
\end{figure}
\begin{figure}[htb]
	\includegraphics[width=1\linewidth]{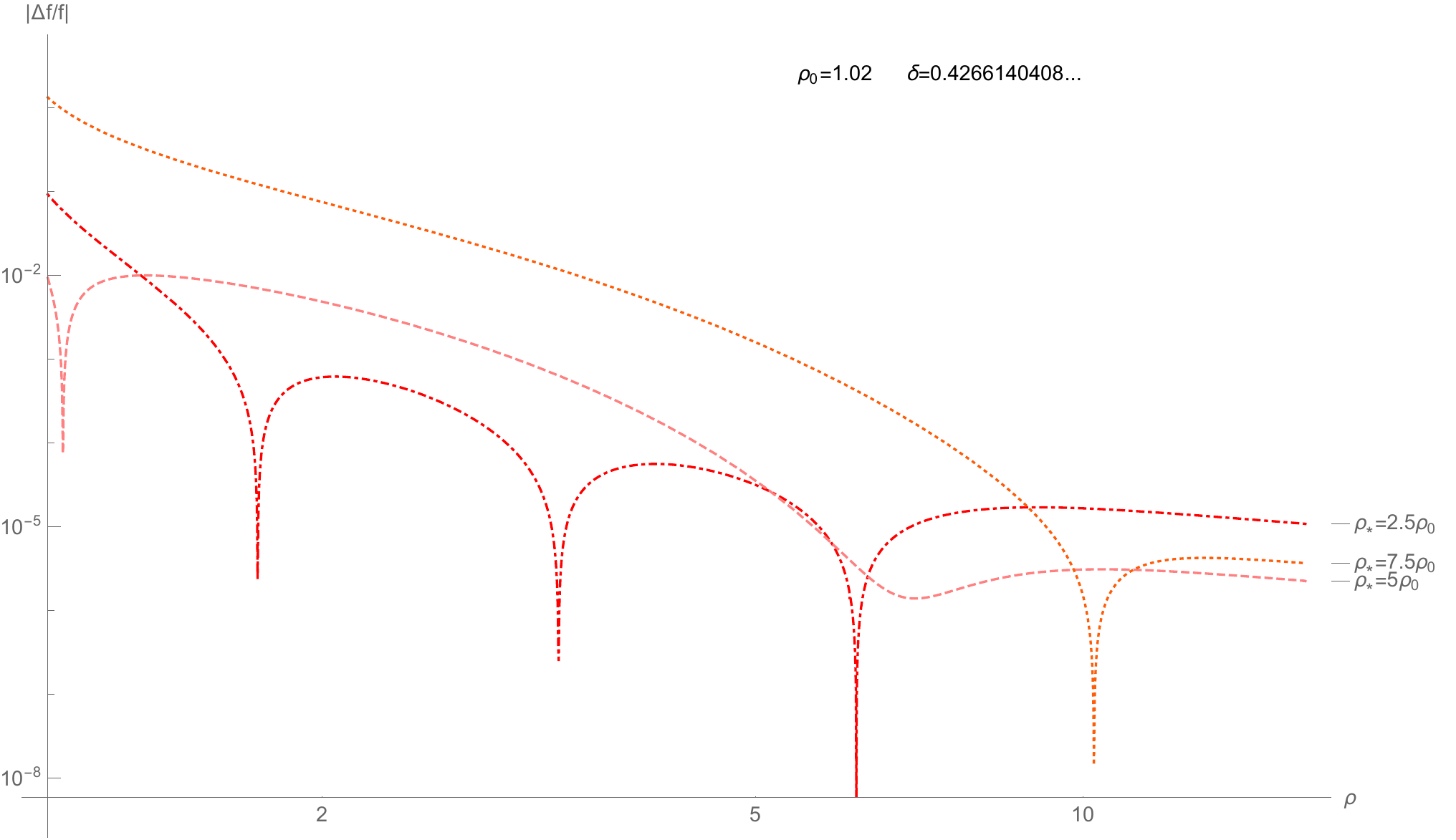}
	\caption{Log-log plots of ${|\Delta f|}/{f}$ against $\rho$ for the third order asymptotic expansion fitted over the range, $\rho_*$ to  
	$\rho_f=40\rho_0$, for various values of $\rho_*$. 	The fitted values for $\lambda$ where, $\lambda_{\rho_*=2.5}=0.79\ldots$(red dot-dashed line), $\lambda_{\rho_*=5.0}=0.70\ldots$(pink dashed line) and $\lambda_{\rho_*=7.5}=0.45\ldots$(orange dotted line).
	}
	\label{fig:log2}
\end{figure}
\Cref{fig:log1,fig:log2} are a log-log plots of 
%%%%%%%%%%
\begin{equation}
\label{plot def}
\frac{|\Delta f|}{\tilde{f}}=\frac{|\tilde{f}-f_\text{Asymptotic}|}{\tilde{f}}~,
\end{equation}
where 
%%%%%%%%%%
\begin{equation}
\label{fN_def}
\tilde{f}=\left\{ 
\begin{array}{l}
\text{Near horizon expansion for $f(\rho)$ for }\rho_0\leqslant\rho\leqslant\rho_0+\tfrac{1}{10}\\
\text{Numerical integration of $f(\rho)$ for } \rho \geqslant \rho_0+\tfrac{1}{10}
\end{array}
\right. ~.
\end{equation}
%%%%%%%%%
The plot in \autoref{fig:log1}  has the same $\delta$ as \autoref{fig:pert1b} but it makes the difference between our fits and the numerical result more visible. The dotted vertical line in the plot corresponds to the point at which we set the boundary conditions for the numerical integration using the near horizon result. It is clear that the expansions are good fits for the numerical data and that increasing the order of the expansion improves the fit. In \autoref{fig:log2},	 we wanted to check the effect of changing the range of $\rho$ over which we fit the asymptotic expansion. We shifted the starting point, $\rho_*$, for the fit, keeping the end point fixed. In this case, the best fit was obtained at $\rho_*=5\rho_0$. The different values we took for $\rho_*$ had very little effect on the best fit for $M$ (and $h_\infty$) but $\lambda$ changed considerably. Ideally, one should try to optimise the value of $\rho_*$ to obtain the best fit since, a priori, one does not know at what radius one can really start to trust the asymptotic expansion. We did not bother doing that since it  had little effect on the mass extracted from the solution which is what we were primarily interested in.      
%%%%%%%%%

\subsection{Thermodynamics}

Having investigated how well our asymptotic expansion fits the numerical results, we can happily go ahead and investigate the thermodynamics. Since we found that using the $\mathcal{O}(3)$ expansion did not alter our results for the mass (and $h_\infty$) very much, we just used the expansion up to $\mathcal{O}(2)$ to find the parameters of the solutions -- these results are summarised in \autoref{ap:num}.

\begin{figure}[htb]
	\includegraphics[width=1\linewidth]{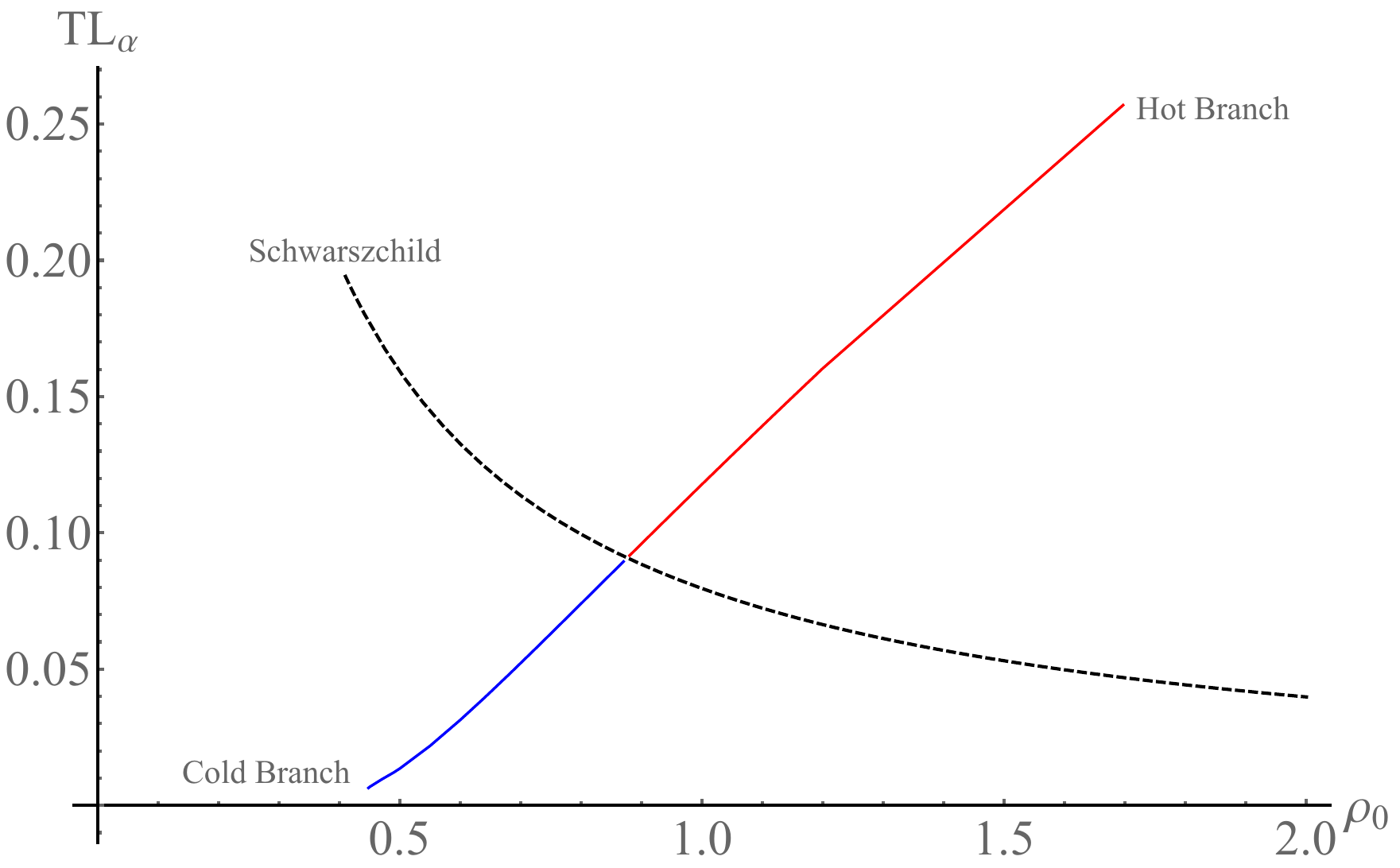}
	\caption{Plot of temperature, $T$, (in units of $1/L_\alpha$) against horizon radius, $\rho_0$, for the Schwarzschild and non-Schwarzschild black holes. The dashed line denotes the Schwarzschild solutions whereas the red and blue solid lines denote the hot and cold non-Schwarzschild branches respectively.   
	}
	\label{fig:Tr0}
\end{figure}
Firstly, we consider \autoref{fig:Tr0} which is a plot of temperature as a function  of $\rho_0$ (recall that $\rho_0$ corresponds to the length scale of the horizon in units of $L_\alpha$). The solid line denotes the non-Schwarzschild solution while the dashed line corresponds to the Schwarzschild one. As mentioned in the introduction, above a certain critical point $\rho_0\approx0.876$ (already found in \cite{Lu:2015cqa}) the non-Schwarzschild solution is hotter than the Schwarzschild one whereas below it the non-Schwarzschild one is colder. We call these two branches the hot (red line) and cold branches (blue line) respectively. Notice that, unlike the Schwarzschild case, the temperature of the non-Schwarzschild decreases as $\rho_0$ decreases. Extrapolating the cold branch to smaller $\rho_0$, it would  appear as if one approaches a zero temperature solution -- unfortunately the numerical ingratiation becomes increasingly unstable and we were not able to investigate this limit.
\begin{figure}[htb]
	\includegraphics[width=1\linewidth]{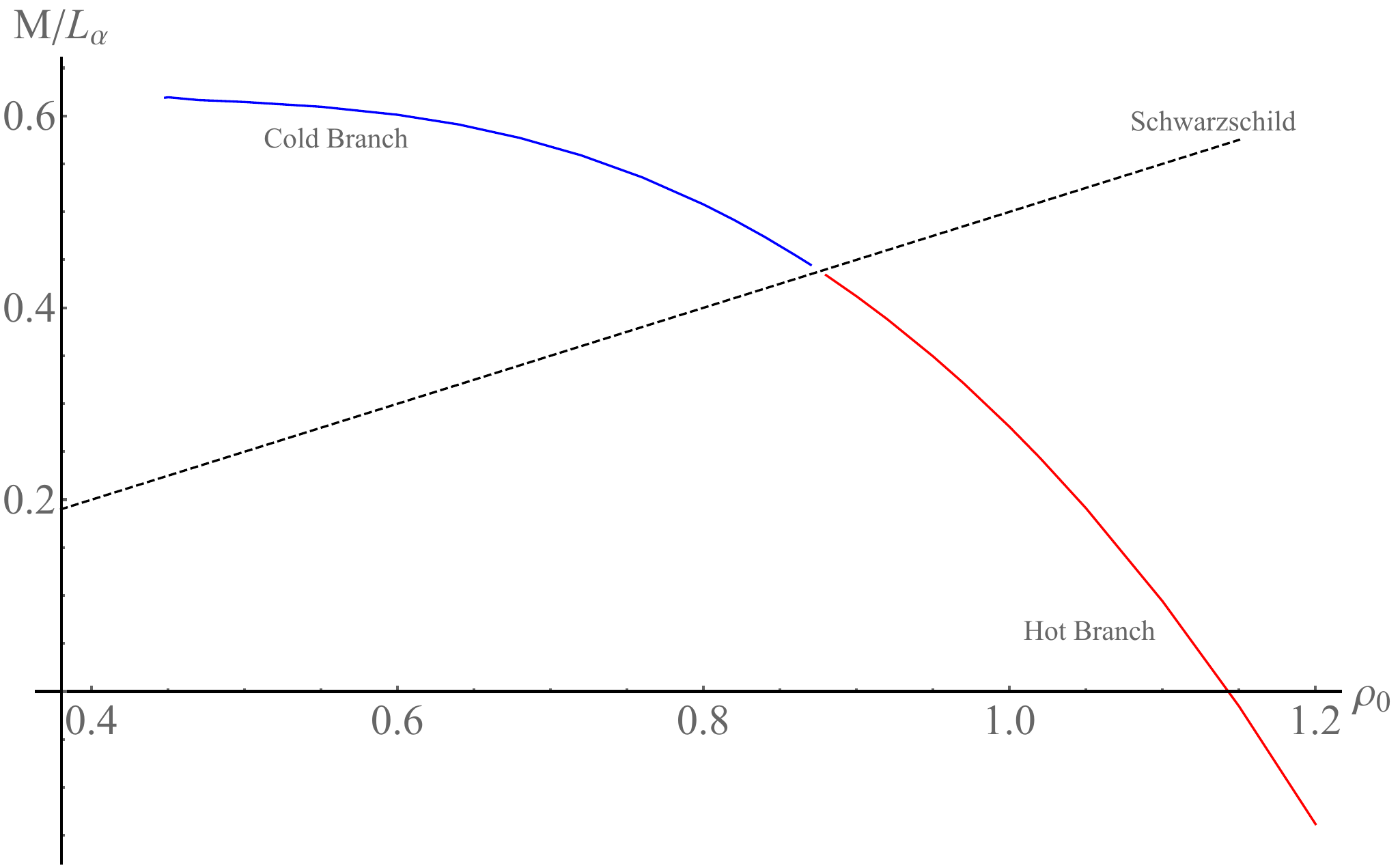}
	\caption{Plot of mass, $M$, (in units of $L_\alpha$) against horizon radius, $\rho_0$, for the Schwarzschild and non-Schwarzschild black holes. The dashed line denotes the Schwarzschild solutions whereas the red and blue solid lines denote the hot and cold non-Schwarzschild branches respectively.
	}
	\label{fig:Mr0}
\end{figure}
Now, looking at \autoref{fig:Mr0}, which is a plot of mass as a function of $\rho_0$, we start to see some more peculiar features of these non-Schwarzschild solutions. Once again the   dashed line corresponds to the Schwarzschild  black hole with the hot and cold non-Schwarzschild branches represented by solid red and blue lines respectively. In contrast to the Schwarzschild solution, the mass of the non-Schwarzschild  solutions increases as $\rho_0$ decreases. On the cold branch the mass seems to saturate at around $0.62 L_\alpha$ as the temperature approaches zero. On the hot branch, the mass actually becomes negative at some point, continuing to become smaller as $\rho_0$ increases without any apparent lower bound. For Schwarzschild black holes, a negative mass is associated with a naked singularity but the negative mass non-Schwarzschild do not seem to violate cosmic censorship.   
\begin{figure}[htb]
	\includegraphics[width=1\linewidth]{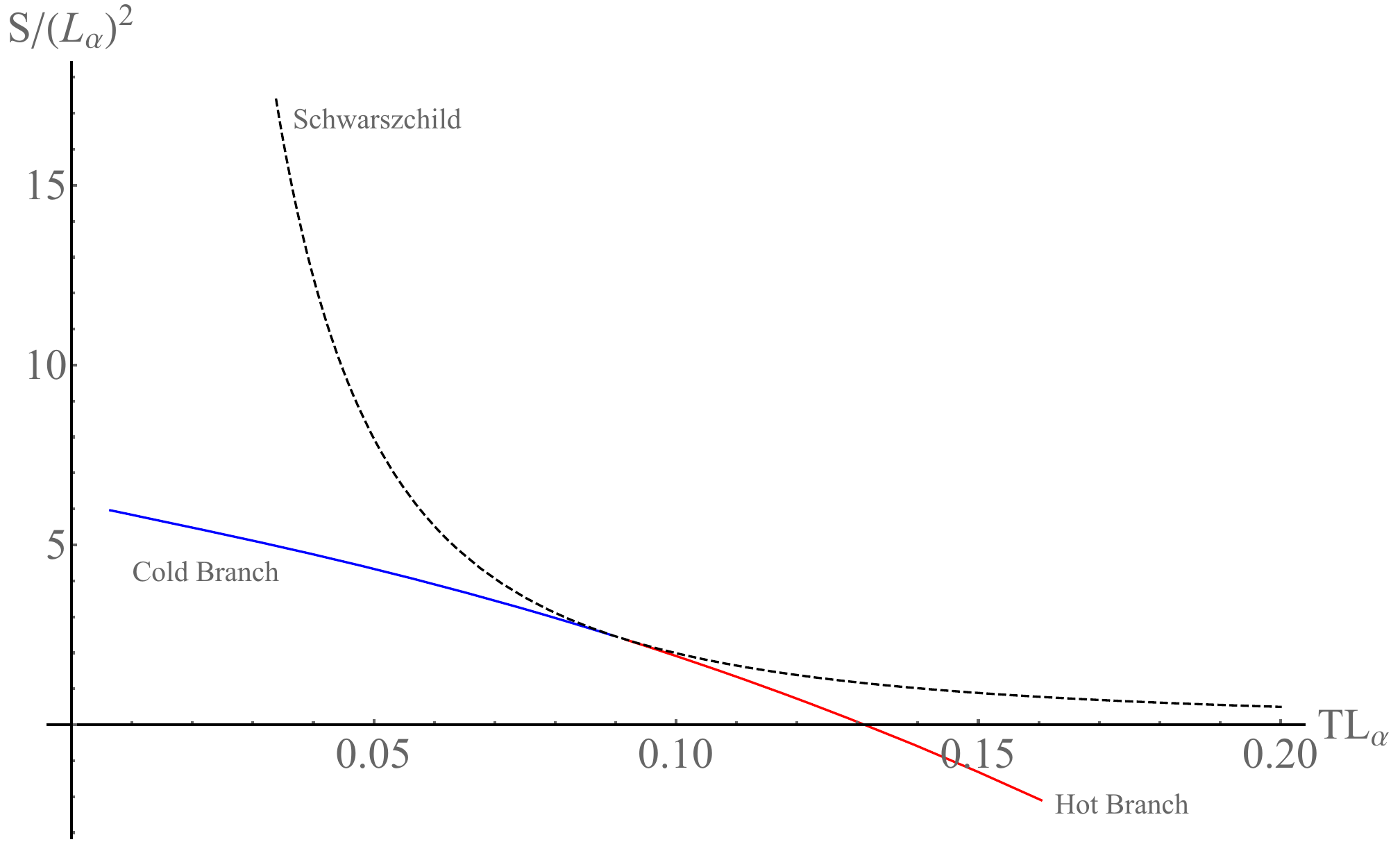}
	\caption{Plot of Wald entropy, $S$, (in units of $L_\alpha^2$) against temperature, $T$, (in units of $1/L_\alpha$), for the Schwarzschild and non-Schwarzschild black holes. The dashed line denotes the Schwarzschild solutions whereas the red and blue solid lines denote the hot and cold non-Schwarzschild branches respectively.
	}
	\label{fig:ST}
\end{figure}

While  negative mass solutions may be a bit troubling, \autoref{fig:ST}, which is a plot of entropy as a function of temperature, shows a severe pathology of the hot branch. Like the Schwarzschild solution, the entropy of the non-Schwarzschild solutions decreases as the temperature increases but there are important differences. The entropy of the Schwarzschild black hole decreases as they become small and hot, with the entropy approaching zero as the temperature approaches infinity and the size approaches zero. On the other hand, the entropy of the hot branch of the Schwarzschild solution, continues to decrease below zero as the temperature and size of the black hole increase. A negative entropy, which one would ultimately like to identify with the logarithm of the number of microstates, does not make any sense. Bizarre features aside, we note that on the cold branch, the entropy seems to approach a finite value as we approach zero temperature and furthermore the non-Schwarzschild solutions always seem to have lower entropy than the Schwarzschild ones.       

\begin{figure}[htb]
	\includegraphics[width=1\linewidth]{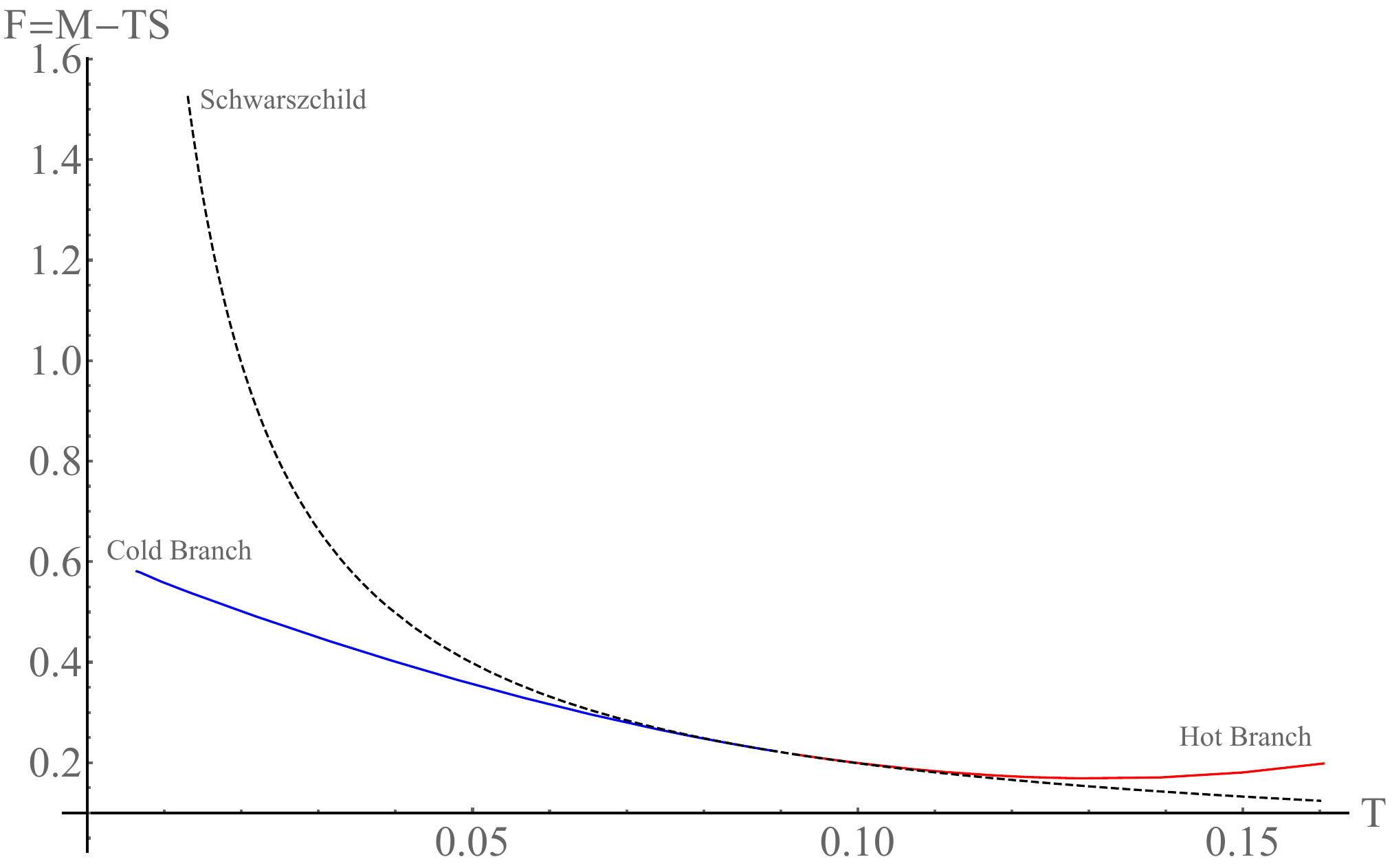}
	\caption{Plot of free energy, $F=M-TS$, against temperature, $T$, (in units $L_\alpha=1$), for the Schwarzschild and non-Schwarzschild black holes. The dashed line denotes the Schwarzschild solutions whereas the red and blue solid lines denote the hot and cold non-Schwarzschild branches respectively.
	}
	\label{fig:FT}
\end{figure}
Finally, in \autoref{fig:FT}, we plot the free energy, $F=M-TS$, of our solutions. Notably, the free energy of the cold branch is lower than corresponding Schwarzschild black hole with the same temperature where as the hot branch has a higher free energy than its Schwarzschild counter part. Like the Schwarzschild black hole, the free energy of the cold branch increases as one decreases the temperature. Unlike the Schwarzschild solution, it appears to tend to a finite value as one approaches zero temperature. Unlike the Schwarzschild black hole, whose free energy decreases monotonically, the hot branch's free energy initially decreases, but then starts increasing, with increasing temperature.  

\section{Conclusions and Outlook}\label{sec:conc}

We found convincing further evidence for the existence of non-Schwarzschild black holes as mathematical solutions to \eqref{eq:Action}. We showed that the numerical solutions matched well with our asymptotic expansion and that the expansions seemed to work well at large distances from the horizon. Unfortunately
this class of black holes does not seem to be physically reasonable. The fact that only small (on the length scale of $L_\alpha$) solutions seem to exist\footnote{One can construct large solutions with negative entropy but these are clearly unphysical.}, invalidates our original assumptions about treating gravity as some effective theory and neglecting even higher order corrections. Truncating \eqref{eq:Action} at $\mathcal{O}\left({L_P^2}/{L_C^2}\right)$, and quantising, leads to ghosts \cite{stelle1978classical}. Perhaps, these ghosts are responsible for some of the strange features, like negative entropy and mass, of the hot branch of non-Schwarzschild black holes.    

 It would appear, that in the context of effective theory,  there  is no higher derivative hair at $\mathcal{O}\left({L_P^2}/{L_C^2}\right)$, in asymptotically flat space -- hence the title of our paper.  We did try to search for other branches of non-Schwarzschild solutions with more physically reasonable solutions by scanning over a large range of $\delta$'s for a given $\rho_0$ . Although we did not succeed, we may just have missed them just as the cold branch was missed in \cite{Lu:2015cqa} -- further study would be required to convincingly exclude this possibility.
 
 Recently, neglecting the  $\mathcal{O}\left({L_P^2}/{L_C^2}\right)$  order terms and going to the next order,  \cite{Hennigar:2016gkm, Bueno:2016lrh,Hennigar:2017ego} have constructed black holes in pure gravity. These solutions seem to be more reasonable than the ones discussed in this paper. We hope that generalising our results to higher dimensions and asymptotically AdS space-times will lead to new non-pathological solutions without having to resort to complexity of the next order. 
 In particular adding a cosmological constant, and working in asymptotically AdS space, it was shown in \cite{Lu:2015cqa}, that one necessarily has $R=\Lambda$ everywhere. This  means that one can not neglect terms in the equations of motion  (c.f. \eqref{eq:eom}) involving $\beta$ as we did in flat space, this may lead to a richer solution space in AdS.

\section*{Acknowledgements} 

The work of KG is supported in part by the South African National Research Foundation grant \# 76970. JJM is supported by a NiTheP MSc bursary.
KG would like to acknowledge useful discussions with Robert de Mello Koch, Vishnu Jejjala and  Dieter L{\"u}st.

\appendix

\section{Asymtotic expansion}
\label{ap:asymp}

In this appendix we present some of the calculations and results we obtained for the asymptotic expansion at  $3^\text{rd}$ order.
Plugging in the second order result and expanding the relevant equations of motion to  third order we find the following source terms: 
\begin{eqnarray}
\label{eq:J3}
J_{(3)}&=& c_1^3 \left(\tfrac{2}{\rho^{6}}\right) \left(\rho ^2-20\right)\cr
&&+c_1^2c_2e^{-\rho}\left(\tfrac{1}{4\rho^{6}}\right)
\left\{
  \rho e^{2 \rho } \text{Ei}(-2 \rho ) \left(2\rho ^4-10 \rho ^3+24 \rho ^2-42 \rho +42\right) +14 \rho ^4+77 \rho ^3 \right.\cr
  &&\phantom{+c_1^2c_2e^{-\rho}\left(\tfrac{1}{4\rho^{6}}\right)}
	  \left.\;+254 \rho ^2+371 \rho +380 
   -2 \left(\rho ^4+5 \rho ^3+12 \rho ^2+21 \rho +21\right) \rho  \log
     (\rho )\right\}\cr
     &&+c_1c_2^2e^{-2\rho}\left(\tfrac{1}{64\rho^{6}}\right)
     \left\{
     -2080 \rho ^5-5778 \rho ^4-10196 \rho ^3-12438 \rho ^2-7348 \rho -3200
     \right.\cr
     &&\phantom{+c_1c_2^2e^{-2\rho}\left(\tfrac{1}{64\rho^{6}}\right)}
     \;-69 \left(\rho ^4-5 \rho ^3+12 \rho ^2-21 \rho +21\right)  \rho  e^{3 \rho }\text{Ei}(-3 \rho )\cr
     &&\phantom{+c_1c_2^2e^{-2\rho}\left(\tfrac{1}{64\rho^{6}}\right)}
     \;+16   \left(\rho ^4-17 \rho ^2-27\right)  \rho e^{2 \rho }\text{Ei}(-2 \rho )\cr
      &&\phantom{+c_1c_2^2e^{-2\rho}\left(\tfrac{1}{64\rho^{6}}\right)}
     \;+69 \left(\rho ^4+5 \rho ^3+12 \rho ^2+21 \rho +21\right)\rho e^{\rho } \text{Ei}(-\rho ) \cr
      &&\phantom{+c_1c_2^2e^{-2\rho}\left(\tfrac{1}{64\rho^{6}}\right)}
      \left.
      \;+16 \left(30 \rho ^5+51 \rho ^4+70 \rho ^3+71 \rho ^2+54 \rho +27\right) \rho\log(\rho)
      \right\}\cr
      &&+c_2^3e^{-3\rho}\left(\tfrac{1}{128\rho^{6}}\right)
      \left\{
      4992 \rho ^6+13354 \rho ^5+18432 \rho ^4+16690 \rho ^3+8650 \rho ^2+84 \rho -800 
      \right.\cr
      &&\phantom{+c_2^3e^{-3\rho}\left(\tfrac{1}{128\rho^{6}}\right)} 
      \; -69 \left(\rho ^4-17 \rho ^2-27\right) \rho  e^{3 \rho }  \text{Ei}(-3 \rho )\cr
      &&\phantom{+c_2^3e^{-3\rho}\left(\tfrac{1}{128\rho^{6}}\right)}       
      \left.
      \;-69   \left(30 \rho ^5+51 \rho ^4+70 \rho ^3+71 \rho ^2+54 \rho +27\right) \rho e^{\rho } \text{Ei}(-\rho )
      \right\}~,
\end{eqnarray}
and 
\begin{multline}
\label{eq:f3}
F_{(3)}=c_1^3\left(\tfrac{4}{\rho ^3}\right)\\
+ c_1^2c_2\left(\tfrac{e^{-\rho }}{8 \rho ^3}\right) 
\left({ \left(-2 e^{2 \rho } \rho  (\rho  (7 \rho -9)+9) \text{Ei}(-2 \rho )-\rho  (58 \rho +73)+2 \rho  (\rho  (7 \rho +9)+9) \log (\rho )-86\right)}\right)\\
+c_1c_2^2 \left(\tfrac{e^{-2\rho}}{128\rho^3}\right)
\left(128 e^{2 \rho } \rho  \left(2 \rho ^2+3\right) \text{Ei}(-2 \rho )+69 e^{3 \rho } \rho  (\rho  (7 \rho -9)+9) \text{Ei}(-3 \rho )\right)\\
+c_1c_2^2 \left(\tfrac{e^{-2\rho}}{128\rho^3}\right)
\left(2 (\rho  (\rho  (640 \rho +1711)+942)-64 \rho  (\rho  (\rho  (\rho +8)+6)+3) \log (\rho )+512)\right)\\
+c_1c_2^2 \left(\tfrac{e^{-2\rho}}{128\rho^3}\right)
\left(-69 e^{\rho } \rho  (\rho  (7 \rho +9)+9) \text{Ei}(-\rho )\right)\\
+c_2^3 \left(\tfrac{e^{-3\rho}}{32\rho^3}\right)
\left(
96 e^{2 \rho } \rho ^4 \text{Ei}(-2 \rho )-69 e^{3 \rho } \left(2 \rho ^2+3\right) \rho  \text{Ei}(-3 \rho )+69 e^{\rho } (\rho  (\rho  (\rho +8)+6)+3) \rho  \text{Ei}(-\rho )
\right)\\
+c_2^3 \left(\tfrac{e^{-3\rho}}{32\rho^3}\right)
\left(
-(\rho  (\rho  (96
   \rho +631)+615)+114) \rho -22
\right)~,
\end{multline}
These sources need to be plugged into \eqref{eq:ynsol1} and \eqref{eq:ynsol2} and integrated.
Given the fact that $J_{(3)}$ and $F_{(3)}$ are quite large,  we found it more manageable to use the linearity of \eqref{eq:ynsol1}-\eqref{eq:ynsol2},  
 and calculate the pieces, $A_{(3,i)}$ and $h_{(3,i)}$, independently .
For the first term (which is what one would have for the Schwarzschild black hole)  one finds
\begin{eqnarray}
\label{eq:g33}
(M^3 m_2^3) A_{(3,3)}&=& \frac{8 M^3}{r^3}~,\cr
h_{(3,3)}&=&0~,
\end{eqnarray}
as expected (cf. \eqref{eq:Sw_exp1}-\eqref{eq:Sw_exp2}). For the other parts of the metric, we were unable to express all the integrals in terms of known special functions.  In fact we found in easier to just calculate $h'_{(3)}$ rather than just $h_{(3)}$ -- fortunately this is sufficient for matching the numerical data and asymptotic expressions. In fact, since $h$ (as opposed to its derivatives) does not appear in the equations of motion,  \eqref{eq:ynsol1}-\eqref{eq:ynsol2}, one could even proceed to the next order without calculating $h$. For completeness, we present the rest of our results at $3^\text{rd}$ order below:
%%%%%%%%%%
%%%%%%%%%%
\begin{eqnarray}
\label{label}
%%%%%%%%%%%%%%%%%%%%%%
A_{(3,0)}&=& 
\tfrac{4761}{2048}e^{-\rho }  X^{-1,3}(\rho )
\left(1+\tfrac{1}{\rho}\right)
+\tfrac{4761}{1024}e^{\rho }X^{1,3}(\rho )
\left(1-\tfrac{1}{\rho}\right)
\cr
%%%%%%%%%%%%%%%%%%%%%%%%%%%%%%%%%%%%%%%%%%%%
&&+\tfrac{621}{512\rho^2}\left({ e^{-2 \rho } \text{Ei}(-\rho )}-{ \text{Ei}(-3 \rho )}\right)
-\tfrac{4761}{4096}e^{-\rho }\text{Ei}(-\rho )^2\left(1+\tfrac{1}{\rho}\right)
\cr
%%%%%%%%%%%%%%%%%%%%%%%%%%%%%%%%%%%%%%%%%%%%
&&+\tfrac{345}{128} e^{-2 \rho } \text{Ei}(-\rho ) -\tfrac{4761}{2048} \text{Ei}(-3 \rho ) \text{Ei}(-\rho )e^{\rho }\left(1-\tfrac{1}{\rho}\right)
\cr
%%%%%%%%%%%%%%%%%%%%%%%%%%%%%%%%%%%%%%%%%%%%
&&
-\tfrac{565}{256} \text{Ei}(-4 \rho )e^{\rho }\left(1-\tfrac{1}{\rho}\right)
+\tfrac{69}{128}\text{Ei}(-3 \rho )\left(1-\tfrac{1}{\rho}\right)
\cr
%%%%%%%%%%%%%%%%%%%%%%%%%%%%%%%%%%%%%%%%%%%%
&&
-\tfrac{373}{512} e^{-\rho } \text{Ei}(-2 \rho )\left(1+\tfrac{1}{\rho}\right)+\tfrac{1449}{512 \rho } e^{-2 \rho } \text{Ei}(-\rho )
\cr
%%%%%%%%%%%%%%%%%%%%%%%%%%%%%%%%%%%%%%%%%%%%
&&
+\tfrac{5 }{32 \rho ^3}e^{-3 \rho }
-\tfrac{147 }{128 \rho^2}e^{-3 \rho }
-\tfrac{1845 }{512 \rho }e^{-3 \rho }
-\tfrac{13 }{8}e^{-3 \rho }~,\\
%%%%%%%%%%%%%%%%%%%%%%%%%%%%%%%%%%%%%%%%
A_{(3,1)}&=&
\tfrac{207}{128} e^{\rho }
\left(
3 G_{2,3}^{3,0}\left(
\left.
\substack{0,0 \\-1,-1,-1}
\right| 3 \rho 
\right)
- G_{2,3}^{3,0}\left(
\left.
\substack{
1,1 \\
0,0,0}
\right|
3 \rho 
\right)
\right)
-\tfrac{69}{128} e^{-\rho }
\left(
G_{2,3}^{3,0}\left(
\left.
\substack{0,0 \\-1,-1,-1}
\right|
\rho
\right)
+
G_{2,3}^{3,0}\left( \left.
\substack{
	1,1 \\
	0,0,0}
\right|
\rho\right)
\right)
\cr
&& -\tfrac{69}{128} e^{-\rho }X^{-2,3}(\rho )\left(1+\tfrac{1}{\rho}\right)
-\tfrac{69}{64} e^{-\rho } X^{-1,2}(\rho)\left(1+\tfrac{1}{\rho}\right)
-\tfrac{207}{128} e^{2\rho } X^{1,2}(\rho )\left(1-\tfrac{1}{\rho}\right)\cr
&&
+\tfrac{483 }{256 \rho ^2} e^{\rho }\left(\text{Ei}(-3 \rho )-e^{-2\rho } \text{Ei}(-\rho )\right)
+\tfrac{9 }{16 \rho ^2}\text{Ei}(-2 \rho )
+\tfrac{675}{256} e^{\rho } \text{Ei}(-3 \rho )
-\tfrac{1089 }{256 \rho }e^{\rho } \text{Ei}(-3 \rho )\cr
&& -\tfrac{1}{4}\text{Ei}(-2 \rho)
+\tfrac{225}{256} e^{-\rho } \text{Ei}(-\rho )
+\tfrac{69}{128}e^{\rho } \text{Ei}(-2 \rho ) \text{Ei}(-\rho )\left(1-\tfrac{1}{\rho}\right)\cr
&& -\tfrac{189 }{256 \rho }e^{-\rho } \text{Ei}(-\rho )
+\tfrac{69}{64} \text{Ei}(-3\rho ) \log (\rho )e^{\rho }\left(1-\tfrac{1}{\rho}\right)
+\tfrac{69}{64} e^{-\rho }\text{Ei}(-\rho ) \log (\rho)\left(1+\tfrac{1}{\rho}\right)\cr
&& 
+\tfrac{5 }{4 \rho ^3}e^{-2 \rho }+\tfrac{239}{64 \rho ^2} e^{-2 \rho }
-\tfrac{9}{16 \rho ^2}e^{-2 \rho } \log (\rho )
+\tfrac{609 }{128 \rho }e^{-2 \rho }-\tfrac{5}{4} e^{-2 \rho } \log (\rho )
-\tfrac{21 }{16 \rho }e^{-2 \rho } \log(\rho )\\
%%%%%%%%%%%%%%%%%%%%%%%%%%%%%%%%%%%%
%%%%%%%%%%%%%%%%%%%%%%%%%%%%%%%%%%%%
A_{(3,2)}&=&
-e^{\rho } G_{2,3}^{3,0}\left( \left.
\begin{smallmatrix}
0,0 \\
-1,-1,-1 \\
\end{smallmatrix}
\right|2 \rho\right)
+\tfrac{1}{2} e^{\rho } G_{2,3}^{3,0}\left(2 \rho \left|
\begin{smallmatrix}
1,1 \\
0,0,0 \\
\end{smallmatrix}
\right.\right)  
+\tfrac{69 }{1024}
e^{-\rho } G_{2,3}^{3,1}\left( \left.
\begin{smallmatrix}
-1,0 \\
-1,-1,-1 \\
\end{smallmatrix}
\right|\rho\right)\cr
%%%%%%%%%%%%%%%%%%%%%%%%%%%%%%%%%%%%%
&&+\tfrac{1}{2} e^{-\rho } G_{2,3}^{3,1}\left( \left.
\begin{smallmatrix}
-1,0 \\
-1,-1,-1 \\
\end{smallmatrix}
\right|2 \rho
\right)
+\tfrac{69 }{1024}
e^{-\rho } G_{2,3}^{3,1}\left(\rho \left|
\begin{smallmatrix}
0,1 \\
0,0,0 \\
\end{smallmatrix}
\right.\right)
+\tfrac{1}{4} e^{-\rho } G_{2,3}^{3,1}\left(2 \rho \left|
\begin{smallmatrix}
0,1 \\
0,0,0 \\
\end{smallmatrix}
\right.\right)
\cr
&& -\tfrac{69}{2048}e^{\rho }\left(1-\tfrac{1}{\rho}\right) X^{3,-1}(\rho )
-\tfrac{9 }{512 \rho ^2}\text{Ei}(-\rho )
-\tfrac{19 e^{\rho } }{16 \rho ^2}\text{Ei}(-2 \rho )
+\tfrac{9}{512 \rho ^2} e^{-2 \rho }\text{Ei}(\rho )
\cr
&&+\tfrac{69}{4096}\text{Ei}(-\rho )^2e^{\rho }\left(1-\tfrac{1}{\rho}\right)
-\tfrac{69 }{2048} e^{-\rho }\text{Ei}(\rho ) \text{Ei}(-\rho )\left(1+\tfrac{1}{\rho}\right)
+\tfrac{15 }{128 \rho }\text{Ei}(-\rho )
\cr
%%%%%%%%%%%%%%%%%%%%%%%%%
&&
+\tfrac{1}{128}\text{Ei}(-\rho )
-\tfrac{1523}{1536} e^{\rho } \text{Ei}(-2 \rho )
+\tfrac{2483 }{1536\rho }e^{\rho } \text{Ei}(-2 \rho )
+\tfrac{5}{128} e^{-2 \rho } \text{Ei}(\rho )
+\tfrac{21 }{512 \rho }e^{-2 \rho } \text{Ei}(\rho )\cr
%%%%%%%%%%%%%%%%%%%%%%%%%%%%%%%%
&&
-\tfrac{1}{4} e^{\rho } \text{Ei}(-2 \rho ) \log (\rho )
+\tfrac{1}{4 \rho}e^{\rho } \text{Ei}(-2 \rho ) \log (\rho )
-\tfrac{129 }{32 \rho ^3}e^{-\rho }
-\tfrac{95}{64 \rho ^2} e^{-\rho }
+\tfrac{19 }{16 \rho ^2}e^{-\rho } \log (\rho )\cr
%%%%%%%%%%%%%%%%%%%%%%%%%%%%%
&&
-\tfrac{1681 }{1536 \rho }e^{-\rho }
-\tfrac{1}{8} e^{-\rho } \log ^2(\rho )
-\tfrac{1}{8 \rho }e^{-\rho } \log^2(\rho )
+\tfrac{5}{8 \rho } e^{-\rho } \log (\rho )\cr
%%%%%%%%%%%%%%%%%%%%%%%%%%%%%%%%
h'_{(3,0)}&=& 
\tfrac{4761  }{1024 \rho}e^{-\rho }X^{-1,3}(\rho )\left(1+\tfrac{1}{\rho}\right)
+\tfrac{4761}{512 \rho}  X^{1,3}(\rho )e^{\rho }\left(1-\tfrac{1}{\rho}\right)\cr
%%%%%%%%%%%%%%%%%%
&&
-\tfrac{1035}{256 \rho ^3}e^{-2 \rho} \text{Ei}(-\rho )
+\tfrac{1035 }{256 \rho ^3}\text{Ei}(-3 \rho )
-\tfrac{4761}{2048 \rho ^2} e^{-\rho } \text{Ei}(-\rho )^2
+\tfrac{4761 }{1024 \rho^2}e^{\rho } \text{Ei}(-3 \rho ) \text{Ei}(-\rho )\cr
%%%%%%%%%%%%%%%%%%%%%
&&
-\tfrac{897}{256 \rho ^2} e^{-2 \rho } \text{Ei}(-\rho )
+\tfrac{565 }{128 \rho ^2}e^{\rho } \text{Ei}(-4 \rho )
-\tfrac{69}{128 \rho ^2} \text{Ei}(-3 \rho )
-\tfrac{1141 }{256 \rho^2}e^{-\rho } \text{Ei}(-2 \rho )\cr
%%%%%%%%%%%%%%%%%%%%%%%%%%
&&
-\tfrac{4761 }{2048 \rho }e^{-\rho } \text{Ei}(-\rho )^2
-\tfrac{4761}{1024 \rho } e^{\rho } \text{Ei}(-3 \rho ) \text{Ei}(-\rho )
+\tfrac{69 }{16 \rho }e^{-2 \rho } \text{Ei}(-\rho )
-\tfrac{565}{128 \rho } e^{\rho }	\text{Ei}(-4 \rho )\cr
%%%%%%%%%%%%%%%%%%
&&
-\tfrac{207 }{64 \rho }\text{Ei}(-3 \rho )
-\tfrac{1141}{256 \rho } e^{-\rho } \text{Ei}(-2 \rho )
+\tfrac{21}{32 \rho ^4} e^{-3 \rho }
+\tfrac{15 }{64 \rho ^3}e^{-3 \rho }
-\tfrac{793 }{256 \rho ^2}e^{-3\rho }
-\tfrac{9 }{4 \rho }e^{-3 \rho }\\
%%%%%%%%%%%%%%%%%%%%%%%%%%%%%%%%%%%%%%%%%%%%%%%%%%%%%%%%
h'_{(3,1)}&=&
\tfrac{1863}{64} e^{\rho } 
G_{2,3}^{3,0}\left( \left.
\begin{smallmatrix}
-1,-1 \\
-2,-2,-2 \\
\end{smallmatrix}
\right|
3 \rho
\right)
-\tfrac{69}{64} e^{-\rho } 
G_{2,3}^{3,0}\left(\left.
\begin{smallmatrix}
-1,-1 \\
-2,-2,-2 \\
\end{smallmatrix}
\right|
\rho 
\right)
\cr
%%%%%%%%%%%%%%
&&
-\tfrac{69}{64} e^{-\rho } G_{2,3}^{3,0}\left( \left.
\begin{smallmatrix}
0,0 \\
-1,-1,-1 \\
\end{smallmatrix}
\right|
\rho
\right)
-\tfrac{621}{64} e^{\rho } G_{2,3}^{3,0}\left( \left.
\begin{smallmatrix}
0,0 \\
-1,-1,-1 \\
\end{smallmatrix}
\right|3 \rho\right)\cr
&&-\tfrac{69}{64 \rho } e^{-\rho } X^{-2,3}(\rho )\left(1+\tfrac{1}{\rho}\right)
-\tfrac{69 }{32 \rho }e^{-\rho } X^{-1,2}(\rho )\left(1+\tfrac{1}{\rho}\right)\cr
%%%%%%%%%%%%%%%%%%%%%%%%%%%%%
&&
-\tfrac{207 }{64 \rho }X^{1,2}(\rho )e^{\rho }\left(1-\tfrac{1}{\rho}\right)
-\tfrac{69}{64 \rho ^3} e^{\rho } \text{Ei}(-3 \rho )-\tfrac{15}{8 \rho ^3} \text{Ei}(-2 \rho )
+\tfrac{69 }{64 \rho ^3}e^{-\rho } \text{Ei}(-\rho)\cr
%%%%%%%%%%%%%%%%%%%%%%%%%%
&&
-\tfrac{303 }{64 \rho ^2}e^{\rho } \text{Ei}(-3 \rho )
+\tfrac{3}{2 \rho ^2}\text{Ei}(-2 \rho )
-\tfrac{69 }{64 \rho ^2}e^{\rho } \text{Ei}(-2 \rho ) \text{Ei}(-\rho )
+\tfrac{147 }{64 \rho ^2}e^{-\rho }\text{Ei}(-\rho )\cr
%%%%%%%%%%%%%%%%
&&
-\tfrac{69}{32 \rho ^2}e^{\rho } \text{Ei}(-3 \rho ) \log (\rho )
+\tfrac{69}{32 \rho ^2} e^{-\rho } \text{Ei}(-\rho ) \log (\rho )
+\tfrac{675 }{128 \rho}e^{\rho } \text{Ei}(-3 \rho )\cr
%%%%%%%%%%%%%%%%%%%%%%%%%%%%%%%%%%
&&
+\tfrac{3 }{2 \rho }\text{Ei}(-2 \rho )
+\tfrac{69 }{64 \rho }e^{\rho } \text{Ei}(-2 \rho ) \text{Ei}(-\rho )
+\tfrac{225}{128 \rho } e^{-\rho } \text{Ei}(-\rho )
+\tfrac{69}{32 \rho } e^{\rho } \text{Ei}(-3 \rho ) \log (\rho)\cr
%%%%%%%%%%%%%%%%%%
&&
+\tfrac{69}{32 \rho } e^{-\rho } \text{Ei}(-\rho ) \log (\rho )
-\tfrac{3}{2 \rho ^4} e^{-2 \rho }
+\tfrac{3 }{4 \rho ^3} e^{-2 \rho }
+\tfrac{15}{8 \rho ^3}  e^{-2 \rho } \log (\rho ) \cr
%%%%%%%%%%%%%%%%%%%%%%%%%%%%%%
&&
+\tfrac{369 }{64 \rho ^2}e^{-2 \rho}
+\tfrac{13 }{8 \rho ^2}e^{-2 \rho } \log (\rho )
-\tfrac{2 }{\rho }e^{-2 \rho } \log (\rho )\cr
h'_{(3,2)}&=&
2 e^{\rho } G_{2,3}^{3,0}\left( \left.
\begin{smallmatrix}
0,0 \\
-1,-1,-1 \\
\end{smallmatrix}
\right|2 \rho\right)
-4 e^{\rho } G_{2,3}^{3,0}\left( \left.
\begin{smallmatrix}
-1,-1 \\
-2,-2,-2 \\
\end{smallmatrix}
\right|2 \rho\right)
+2 e^{-\rho } G_{2,3}^{3,1}\left( \left.
\begin{smallmatrix}
-2,-1 \\
-2,-2,-2 \\
\end{smallmatrix}
\right|2 \rho\right)\cr 
%%%%%%%%%%%%%%%%%%
&&
+\tfrac{69}{512} e^{-\rho } G_{2,3}^{3,1}\left( \left.
\begin{smallmatrix}
-2,-1 \\
-2,-2,-2 \\
\end{smallmatrix}
\right|\rho\right)
+\tfrac{69}{512} e^{-\rho } G_{2,3}^{3,1}\left( \left.
\begin{smallmatrix}
-1,0 \\
-1,-1,-1 \\
\end{smallmatrix}
\right|\rho\right)
+e^{-\rho } G_{2,3}^{3,1}\left( \left.
\begin{smallmatrix}
-1,0 \\
-1,-1,-1 \\
\end{smallmatrix}
\right|2 \rho\right)\cr
%%%%%%%%%%%%%%%%%%%%%%%%%%%%%%
&&
-\tfrac{69 }{1024 \rho }X^{3,-1}(\rho )e^{\rho }\left(1-\tfrac{1}{\rho}\right) 
-\tfrac{9 }{256 \rho ^3}\text{Ei}(-\rho )
-\tfrac{1}{8 \rho ^3}e^{\rho } \text{Ei}(-2 \rho )
+\tfrac{9 }{256 \rho ^3}e^{-2 \rho} \text{Ei}(\rho )\cr
%%%%%%%%%%%%%%%%%%%%%%%%%%%%%%%%%%
&&
-\tfrac{69}{2048 \rho ^2} e^{\rho } \text{Ei}(-\rho )^2
-\tfrac{69 }{1024 \rho ^2} e^{-\rho } \text{Ei}(\rho ) \text{Ei}(-\rho )
+\tfrac{15 }{64 \rho ^2}\text{Ei}(-\rho )
+\tfrac{1139}{768 \rho ^2}e^{\rho } \text{Ei}(-2 \rho )\cr
%%%%%%%%%%%%%%%%%%%%%%%%%%%%%%%%%%%%%%
&&
+\tfrac{21 }{256 \rho ^2}e^{-2 \rho } \text{Ei}(\rho )
+\tfrac{1}{2 \rho ^2}e^{\rho } \text{Ei}(-2 \rho ) \log (\rho )
+\tfrac{69}{2048 \rho} e^{\rho } \text{Ei}(-\rho )^2
-\tfrac{69}{1024 \rho } e^{-\rho } \text{Ei}(\rho ) \text{Ei}(-\rho )\cr
&&
+\tfrac{1}{64 \rho }\text{Ei}(-\rho )
-\tfrac{1523}{768 \rho } e^{\rho } \text{Ei}(-2 \rho )
+\tfrac{5 }{64 \rho}e^{-2 \rho } \text{Ei}(\rho )
-\tfrac{1}{2 \rho }e^{\rho } \text{Ei}(-2 \rho ) \log (\rho )
-\tfrac{43}{16 \rho ^4} e^{-\rho }
+\tfrac{25 }{32 \rho ^3}e^{-\rho }\cr
%%%%%%%%%%%%%%%%%%%%%%%%%%%%%%%
&&
+\tfrac{1}{8 \rho ^3}e^{-\rho } \log (\rho )
-\tfrac{1681}{768 \rho^2} e^{-\rho }
-\tfrac{1}{4 \rho ^2}e^{-\rho } \log ^2(\rho )
-\tfrac{1}{2 \rho ^2}e^{-\rho } \log (\rho )
-\tfrac{1}{4 \rho }e^{-\rho } \log ^2(\rho )
%%%%%%%%%%%%%%%%%%%%%%%%%%%%%%%%%%%%%%%%
\end{eqnarray}
%%%%%%%%%%
where $G_{p,q}^{\,m,n} \!\left( \left. \begin{smallmatrix} a_1, \dots, a_p \\ b_1, \dots, b_q \end{smallmatrix} \; \right| \, z \right)$
is the Meijer G-function (see \cite{gradshteyn2014} for a definition) and the function $X$ is short hand for the only integral we could not easily evaluate:
%%%%%%%%%%
\begin{equation}
\label{eq:Xdef2}
X^{a,b}(\rho)=\int^\rho_1 du \frac{e^{-au}{\text{Ei}(-b u)}}{u}-X^{a,b}~.
\end{equation}
where $X^{a,b}$ are integration constants. 
Comparing the form of \eqref{eq:A1sol} with the terms multiplying $X$ in the expression for $A_{3}$ and $h_{3}$, we see that the constants $X^{a,b}$ can be absorbed into the constants arising at first order, namely $c_1,c_2$ and $c_3$. This effectively means that the  constants $X^{a,b}$ can be chosen so that $X^{a,b}(\rho)\rightarrow 0$ as $\rho\rightarrow\infty$ -- we numerically calculated the required constants which are presenting in \autoref{tab:Xab}.
%%%%%%%%%
\begin{table}[htb]
	\centering
	\begin{tabular}{l|l|l}
		$a$ & $b$ & $X^{a,b}$ \\
		\hline 
		$-$2&3& $-$0.04105214790548301\ldots \\
		$-$1&2 &$-$0.05747976304091013\ldots\\
		$-$1&3& $-$0.010232258484269736\ldots\\
		1&2& $-$0.004004262372543252\ldots\\
		1&3& $-$0.0008608700400870111\ldots\\
		2&3& $-$0.00026726321183103313\ldots\\
		3 &$-$1 & \phantom{-}0.03496047797048475\ldots\\
	\end{tabular}
	\caption{Numerically calculated values of the constants $X^{a,b}$}	\label{tab:Xab}
\end{table}
%Definition of Meijers G function

%\begin{equation}
%\label{eq:Meijers G}
%G_{p,q}^{\,m,n} \!\left( \left. \begin{matrix} a_1, \dots, a_p \\ b_1, \dots, b_q \end{matrix} \; \right| \, z \right) = \frac{1}{2 \pi i} \int_L \frac{\prod_{j=1}^m \Gamma(b_j - s) \prod_{j=1}^n \Gamma(1 - a_j +s)} {\prod_{j=m+1}^q \Gamma(1 - b_j + s) \prod_{j=n+1}^p \Gamma(a_j - s)} \,z^s \,ds,
%\end{equation}
\newpage
\section{Numerical results}\label{ap:num}

\begin{table}[htb]
\centering
$
\begin{array}{c|c|c|c|c|c|c}
	{M}/{L_\alpha} & \rho_0 & \delta  & \lambda/{L_\alpha}  & h_\infty & S/{(L_\alpha)}^2 & T L_\alpha \\
	\hline
	\text{ 0.619} & \text{ 0.448} & -0.8485292670 & \text{ 0.294} & \text{ 52.200} & \text{ 5.960} & \text{ 0.006} \\
	\text{ 0.619} & \text{ 0.450} & -0.8456315816 & \text{ 0.178} & \text{ 47.400} & \text{ 5.950} & \text{ 0.007} \\
	\text{ 0.616} & \text{ 0.470} & -0.8209536967 & \text{ 2.370} & \text{ 26.500} & \text{ 5.850} & \text{ 0.010} \\
	\text{ 0.615} & \text{ 0.490} & -0.7966829072 & \text{ 2.390} & \text{ 17.900} & \text{ 5.760} & \text{ 0.012} \\
	\text{ 0.614} & \text{ 0.500} & -0.7835746355 & \text{ 2.330} & \text{ 14.900} & \text{ 5.710} & \text{ 0.014} \\
	\text{ 0.609} & \text{ 0.550} & -0.7103444309 & \text{ 1.720} & \text{ 6.980} & \text{ 5.410} & \text{ 0.022} \\
	\text{ 0.601} & \text{ 0.600} & -0.6263338362 & \text{ 1.600} & \text{ 4.010} & \text{ 5.070} & \text{ 0.031} \\
	\text{ 0.591} & \text{ 0.640} & -0.5521551177 & \text{ 1.480} & \text{ 2.840} & \text{ 4.760} & \text{ 0.040} \\
	\text{ 0.577} & \text{ 0.680} & -0.4721128117 & \text{ 1.280} & \text{ 2.140} & \text{ 4.420} & \text{ 0.048} \\
	\text{ 0.559} & \text{ 0.720} & -0.3864156113 & \text{ 1.090} & \text{ 1.690} & \text{ 4.060} & \text{ 0.057} \\
	\text{ 0.536} & \text{ 0.760} & -0.2952111978 & \text{ 0.838} & \text{ 1.380} & \text{ 3.670} & \text{ 0.065} \\
	\text{ 0.508} & \text{ 0.800} & -0.1986058208 & \text{ 0.553} & \text{ 1.160} & \text{ 3.260} & \text{ 0.074} \\
	\text{ 0.491} & \text{ 0.820} & -0.1483029212 & \text{ 0.448} & \text{ 1.070} & \text{ 3.040} & \text{ 0.078} \\
	\text{ 0.474} & \text{ 0.840} & -0.0966772370 & \text{ 0.254} & \text{ 0.992} & \text{ 2.820} & \text{ 0.083} \\
	\text{ 0.455} & \text{ 0.860} & -0.0437354265 & \text{ 0.094} & \text{ 0.924} & \text{ 2.600} & \text{ 0.087} \\
	\text{ 0.445} & \text{ 0.870} & -0.0167728192 & \text{ 0.042} & \text{ 0.894} & \text{ 2.480} & \text{ 0.089} \\
	\hline
	\text{ 0.434} & \text{ 0.880} & \text{ 0.0105358503} & -0.022 & \text{ 0.865} & \text{ 2.370} & \text{ 0.092} \\
	\text{ 0.412} & \text{ 0.900} & \text{ 0.0660769651} & -0.135 & \text{ 0.813} & \text{ 2.130} & \text{ 0.096} \\
	\text{ 0.388} & \text{ 0.920} & \text{ 0.1229337407} & -0.248 & \text{ 0.766} & \text{ 1.890} & \text{ 0.100} \\
	\text{ 0.349} & \text{ 0.950} & \text{ 0.2106509922} & -0.416 & \text{ 0.705} & \text{ 1.510} & \text{ 0.107} \\
	\text{ 0.321} & \text{ 0.970} & \text{ 0.2707449680} & -0.527 & \text{ 0.669} & \text{ 1.250} & \text{ 0.111} \\
	\text{ 0.276} & \text{ 1.000} & \text{ 0.3633022447} & -0.689 & \text{ 0.622} & \text{ 0.859} & \text{ 0.118} \\
	\text{ 0.243} & \text{ 1.020} & \text{ 0.4266143280} & -0.794 & \text{ 0.594} & \text{ 0.588} & \text{ 0.122} \\
	\text{ 0.191} & \text{ 1.050} & \text{ 0.5239879218} & -0.949 & \text{ 0.556} & \text{ 0.171} & \text{ 0.129} \\
	\text{ 0.094} & \text{ 1.100} & \text{ 0.6926768791} & -1.200 & \text{ 0.502} & -0.551 & \text{ 0.139} \\
	-0.015 & \text{ 1.150} & \text{ 0.8693454710} & -1.440 & \text{ 0.458} & -1.310 & \text{ 0.150} \\
	-0.138 & \text{ 1.200} & \text{ 1.0539754470} & -1.690 & \text{ 0.421} & -2.100 & \text{ 0.160} \\
\end{array}
$
\label{tab:results}
\caption{Summary of non-Schwarzschild black holes obtain by numerically matching near horizon data with asymptotic parameters. }
\end{table}

\bibliographystyle{utphys}
\bibliography{asymp}

\providecommand{\href}[2]{#2}\begingroup\raggedright\begin{thebibliography}{10}

\bibitem{jebsen1921general}
J.~Jebsen, ``On the general spherically symmetric solutions of einstein's
  gravitational equations in vacuo.,''
  \href{http://dx.doi.org/10.1007/s10714-005-0168-y}{{\em Arkiv for Matematik,
  Astronomi och Fysik} {\bfseries 15} (1921) }.

\bibitem{birkhoff1923relativity}
G.~D. Birkhoff and R.~E. Langer, {\em Relativity and modern physics}, vol.~1.
\newblock Harvard University Press Cambridge, 1923.

\bibitem{ANDP:ANDP19233771804}
W.~Alexandrow, ``{\"U}ber den kugelsymmetrischen vakuumvorgang in der
  einsteinschen gravitationstheorie,''
  \href{http://dx.doi.org/10.1002/andp.19233771804}{{\em Annalen der Physik}
  {\bfseries 377} no.~18, (1923) 141--152}.
  \url{http://dx.doi.org/10.1002/andp.19233771804}.

\bibitem{eiesland1925group}
J.~Eiesland, ``The group of motions of an einstein space,''
  \href{http://dx.doi.org/10.1090/S0002-9947-1925-1501308-7}{{\em Transactions
  of the American Mathematical Society} {\bfseries 27} no.~2, (1925) 213--245}.

\bibitem{Lu:2015cqa}
H.~L{\"u}, A.~Perkins, C.~N. Pope, and K.~S. Stelle, ``{Black Holes in
  Higher-Derivative Gravity},''
  \href{http://dx.doi.org/10.1103/PhysRevLett.114.171601}{{\em Phys. Rev.
  Lett.} {\bfseries 114} no.~17, (2015) 171601},
\href{http://arxiv.org/abs/1502.01028}{{\ttfamily arXiv:1502.01028 [hep-th]}}.
%%CITATION = ARXIV:1502.01028;%%.

\bibitem{Oliva:2011xu}
J.~Oliva and S.~Ray, ``{Birkhoff's Theorem in Higher Derivative Theories of
  Gravity},'' \href{http://dx.doi.org/10.1088/0264-9381/28/17/175007}{{\em
  Class. Quant. Grav.} {\bfseries 28} (2011) 175007},
\href{http://arxiv.org/abs/1104.1205}{{\ttfamily arXiv:1104.1205 [gr-qc]}}.
%%CITATION = ARXIV:1104.1205;%%.

\bibitem{Oliva:2012zs}
J.~Oliva and S.~Ray, ``{Birkhoff's Theorem in Higher Derivative Theories of
  Gravity II},'' \href{http://dx.doi.org/10.1103/PhysRevD.86.084014}{{\em Phys.
  Rev.} {\bfseries D86} (2012) 084014},
\href{http://arxiv.org/abs/1201.5601}{{\ttfamily arXiv:1201.5601 [gr-qc]}}.
%%CITATION = ARXIV:1201.5601;%%.

\bibitem{bh1}
J.~D. Bekenstein, ``{Black holes and entropy},''
\href{http://dx.doi.org/10.1103/PhysRevD.7.2333}{{\em Phys.Rev.} {\bfseries D7}
  (1973) 2333--2346}.
%%CITATION = PHRVA,D7,2333;%%.

\bibitem{haw}
S.~Hawking, ``{Particle Creation by Black Holes},''
\href{http://dx.doi.org/10.1007/BF02345020}{{\em Commun.Math.Phys.} {\bfseries
  43} (1975) 199--220}.
%%CITATION = CMPHA,43,199;%%.

\bibitem{Wald:1993nt}
R.~M. Wald, ``{Black hole entropy is the Noether charge},''
  \href{http://dx.doi.org/10.1103/PhysRevD.48.R3427}{{\em Phys. Rev.}
  {\bfseries D48} no.~8, (1993) R3427--R3431},
\href{http://arxiv.org/abs/gr-qc/9307038}{{\ttfamily arXiv:gr-qc/9307038
  [gr-qc]}}.
%%CITATION = GR-QC/9307038;%%.

\bibitem{Jacobson:1993vj}
T.~Jacobson, G.~Kang, and R.~C. Myers, ``{On black hole entropy},''
  \href{http://dx.doi.org/10.1103/PhysRevD.49.6587}{{\em Phys. Rev.} {\bfseries
  D49} (1994) 6587--6598},
\href{http://arxiv.org/abs/gr-qc/9312023}{{\ttfamily arXiv:gr-qc/9312023
  [gr-qc]}}.
%%CITATION = GR-QC/9312023;%%.

\bibitem{Iyer:1994ys}
V.~Iyer and R.~M. Wald, ``{Some properties of Noether charge and a proposal for
  dynamical black hole entropy},''
  \href{http://dx.doi.org/10.1103/PhysRevD.50.846}{{\em Phys. Rev.} {\bfseries
  D50} (1994) 846--864},
\href{http://arxiv.org/abs/gr-qc/9403028}{{\ttfamily arXiv:gr-qc/9403028
  [gr-qc]}}.
%%CITATION = GR-QC/9403028;%%.

\bibitem{Jacobson:1994qe}
T.~Jacobson, G.~Kang, and R.~C. Myers, ``{Black hole entropy in higher
  curvature gravity},'' in {\em {Heat Kernels and Quantum Gravity Winnipeg,
  Canada, August 2-6, 1994}}.
\newblock 1994.
\newblock \href{http://arxiv.org/abs/gr-qc/9502009}{{\ttfamily
  arXiv:gr-qc/9502009 [gr-qc]}}.
\newblock
\url{http://alice.cern.ch/format/showfull?sysnb=0196054}.
\newblock
%%CITATION = GR-QC/9502009;%%.

\bibitem{Sen:2007qy}
A.~Sen, ``{Black Hole Entropy Function, Attractors and Precision Counting of
  Microstates},'' \href{http://dx.doi.org/10.1007/s10714-008-0626-4}{{\em Gen.
  Rel. Grav.} {\bfseries 40} (2008) 2249--2431},
\href{http://arxiv.org/abs/0708.1270}{{\ttfamily arXiv:0708.1270 [hep-th]}}.
%%CITATION = ARXIV:0708.1270;%%.

\bibitem{stelle1978classical}
K.~Stelle, ``Classical gravity with higher derivatives,''
  \href{http://dx.doi.org/10.1007/BF00760427}{{\em General Relativity and
  Gravitation} {\bfseries 9} no.~4, (1978) 353--371}.

\bibitem{Lu:2015psa}
H.~L{\"u}, A.~Perkins, C.~N. Pope, and K.~S. Stelle, ``{Spherically Symmetric
  Solutions in Higher-Derivative Gravity},''
  \href{http://dx.doi.org/10.1103/PhysRevD.92.124019}{{\em Phys. Rev.}
  {\bfseries D92} no.~12, (2015) 124019},
\href{http://arxiv.org/abs/1508.00010}{{\ttfamily arXiv:1508.00010 [hep-th]}}.
%%CITATION = ARXIV:1508.00010;%%.

\bibitem{hawkingellis}
S.~W. Hawking and G.~F.~R. Ellis, {\em The large scale structure of
  space-time}, vol.~1.
\newblock Cambridge university press, 1973.

\bibitem{Alvarez-Gaume:2015rwa}
L.~Alvarez-Gaume, A.~Kehagias, C.~Kounnas, D.~L{\"u}st, and A.~Riotto,
  ``{Aspects of Quadratic Gravity},''
  \href{http://dx.doi.org/10.1002/prop.201500100}{{\em Fortsch. Phys.}
  {\bfseries 64} no.~2-3, (2016) 176--189},
\href{http://arxiv.org/abs/1505.07657}{{\ttfamily arXiv:1505.07657 [hep-th]}}.
%%CITATION = ARXIV:1505.07657;%%.

\bibitem{trans}
G.~A. {Edgar}, ``{Transseries for beginners},'' {\em ArXiv e-prints} (Jan.,
  2008) , \href{http://arxiv.org/abs/0801.4877}{{\ttfamily arXiv:0801.4877
  [math.RA]}}.

\bibitem{Hennigar:2016gkm}
R.~A. Hennigar and R.~B. Mann, ``{Black holes in Einsteinian cubic gravity},''
\href{http://arxiv.org/abs/1610.06675}{{\ttfamily arXiv:1610.06675 [hep-th]}}.
%%CITATION = ARXIV:1610.06675;%%.

\bibitem{Bueno:2016lrh}
P.~Bueno and P.~A. Cano, ``{Four-dimensional black holes in Einsteinian cubic
  gravity},'' \href{http://dx.doi.org/10.1103/PhysRevD.94.124051}{{\em Phys.
  Rev.} {\bfseries D94} no.~12, (2016) 124051},
\href{http://arxiv.org/abs/1610.08019}{{\ttfamily arXiv:1610.08019 [hep-th]}}.
%%CITATION = ARXIV:1610.08019;%%.

\bibitem{Hennigar:2017ego}
R.~A. Hennigar, D.~Kubiznak, and R.~B. Mann, ``{Generalized quasi-topological
  gravity},''
\href{http://arxiv.org/abs/1703.01631}{{\ttfamily arXiv:1703.01631 [hep-th]}}.
%%CITATION = ARXIV:1703.01631;%%.

\bibitem{gradshteyn2014}
I.~S. Gradshteyn and I.~M. Ryzhik, {\em Table of integrals, series, and
  products}.
\newblock Academic press, 2014.

\end{thebibliography}\endgroup

\end{document}